\documentclass[%
 reprint,
nofootinbib,
 amsmath,amssymb,
 aps,
]{revtex4-2}
\usepackage[colorlinks = true,urlcolor=blue,linkcolor = blue,citecolor=blue]{hyperref}
\usepackage{graphicx}
\usepackage{dcolumn}
\usepackage{bm}

\usepackage[normalem]{ulem} 
\usepackage{color}
\usepackage{comment}
\usepackage[export]{adjustbox}
\usepackage{soul}
\usepackage{physics}
\usepackage{tabularx}
\usepackage{multirow}
\usepackage{tikz}
\usepackage{amsmath,amsfonts, amssymb, amsthm, dsfont}
\usepackage{environ}
\NewEnviron{eqs}{%
\begin{equation}\begin{split}
    \BODY
\end{split}\end{equation}
}

\definecolor{tyler}{rgb}{1,.2,0}

\theoremstyle{definition}
\newtheorem*{definition}{Definition}

\newcommand{\calE}{\mathcal{E}}

\newcommand{\calL}{\mathcal{L}}

\newcommand{\TE}[1]{{\color{tyler}\footnotesize{(TE) #1}}}

\newcommand{\Amirreza}[1]{{\color{blue}#1}}

\begin{document}

\preprint{}

\title{Spacetime Markov length: a diagnostic for fault tolerance via mixed-state phases}%

\author{Amir-Reza Negari$^{1,2}$}
\email{anegari@pitp.ca}
\author{Tyler D. Ellison$^{1}$}
\author{Timothy H. Hsieh$^{1,2}$}


\affiliation{$^1$Perimeter Institute for Theoretical Physics, Waterloo, Ontario N2L 2Y5, Canada}
\affiliation{$^2$Department of Physics and Astronomy, University of Waterloo, Waterloo, Ontario N2L 3G1, Canada}

\begin{abstract}

We establish a correspondence between the fault-tolerance of local stabilizer codes experiencing measurement and physical errors and the mixed-state phases of decohered resource states in one higher dimension. Drawing from recent developments in mixed-state phases of matter, this motivates a diagnostic of fault-tolerance, which we refer to as the spacetime Markov length. This is a length scale determined by the decay of the (classical) conditional mutual information of repeated syndrome measurement outcomes in spacetime. The diagnostic is independent of the decoder, and its divergence signals the intrinsic breakdown of fault tolerance. As a byproduct, we find that decoherence may be useful for exposing transitions from higher-form symmetry-protected topological phases driven by both incoherent and coherent perturbations.    

\end{abstract}

\maketitle

\noindent \textbf{Introduction:} The threshold theorem \cite{aharonov1999faulttolerantquantumcomputationconstant,Kitaev_2003,Knill_1998,shor1997faulttolerantquantumcomputation} is a cornerstone of quantum fault tolerance, ensuring that quantum computations can be reliably performed if error rates are below a certain threshold.  It is well known that the error threshold, in certain cases, can be related to an order-to-disorder phase transition by mapping the decoding problem to a classical statistical-mechanics model~\mbox{\cite{Dennis_2002, Chubb2021statistical, hauser2024informationdynamicsdecoheredquantum}}. However, given the inherent competition between decoherence and error correction, it is natural to wonder whether the threshold can be viewed instead as a transition at the level of quantum states---between distinct mixed-state phases of matter. There has been significant recent progress in understanding mixed-state phases of matter~\cite{diss1, diss2, hastings2011nonzero, coser2019classification, fan2023diagnostics, chen2023separability, bao2023mixedstatetopologicalordererrorfield, sang2023mixed, sang2024stabilitymixedstatequantumphases, lee2023quantum, zou2023channeling, de2022symmetry, ma2023average, zhang2022strange, ma2023topological, rakovszky2024defining,  lu2023mixed, lee2022decoding, zhu2022nishimoris, chen2023symmetryenforced, lessa2024mixedstate, chen2024unconventional, lee2022symmetry, lee2024exact, sohal2024noisy, ellison2024towards, wang2023intrinsic, wang2023topologically, wang2024anomaly, xue2024tensor, guo2024locally, ma2024symmetry,zhang2024quantumcommunicationmixedstateorder,zhang2024strongtoweakspontaneousbreaking1form, lu2024bilayerconstructionmixedstate}, so such a connection would allow for renewed perspectives on quantum error correction enabled by our developing understanding of quantum phases of matter.

A fundamental component of fault tolerance is quantum error correction, which often requires repeating error syndrome measurements to compensate for the fact that the measurements themselves may be faulty. In the idealistic case of perfect measurements, there has been remarkable progress in relating noisy error-correcting codes to mixed-state entanglement properties and mixed-state phases of matter~\cite{fan2023diagnostics,bao2023mixedstatetopologicalordererrorfield, chen2023separability,sang2024stabilitymixedstatequantumphases,sang2023mixed}. The paradigmatic example of this is the toric code subjected to local Pauli noise, in which the breakdown of the quantum memory can be understood as a mixed-state phase transition induced by the noise. Note that, in this context, a mixed-state phase is taken to be an equivalence class of states, where two states are considered to be in the same mixed-state phase if there exist local Lindbladian evolutions that transform one into the other~\cite{coser2019classification}.  

Despite this progress, such a relation for the fully fault-tolerant setting, which includes repeated faulty measurements, has yet to be established. This necessitates taking a spacetime perspective on fault tolerance, in which physical and measurement errors appear at locations throughout spacetime~\cite{gottesman2022opportunitieschallengesfaulttolerantquantum}. This perspective is manifest in measurement-based quantum computation (MBQC), where the time dimension of a quantum error-correcting process is replaced with an extra spatial dimension~\mbox{\cite{Raussendorf2001oneway, Bolt_2016, Brown_2020}}. More specifically, the spacetime quantum circuit on a system with $D$ spatial dimentions is realized by performing measurements on a resource state in $D+1$ spatial dimensions. 

\begin{figure}[t]
\includegraphics[width=0.48\textwidth]{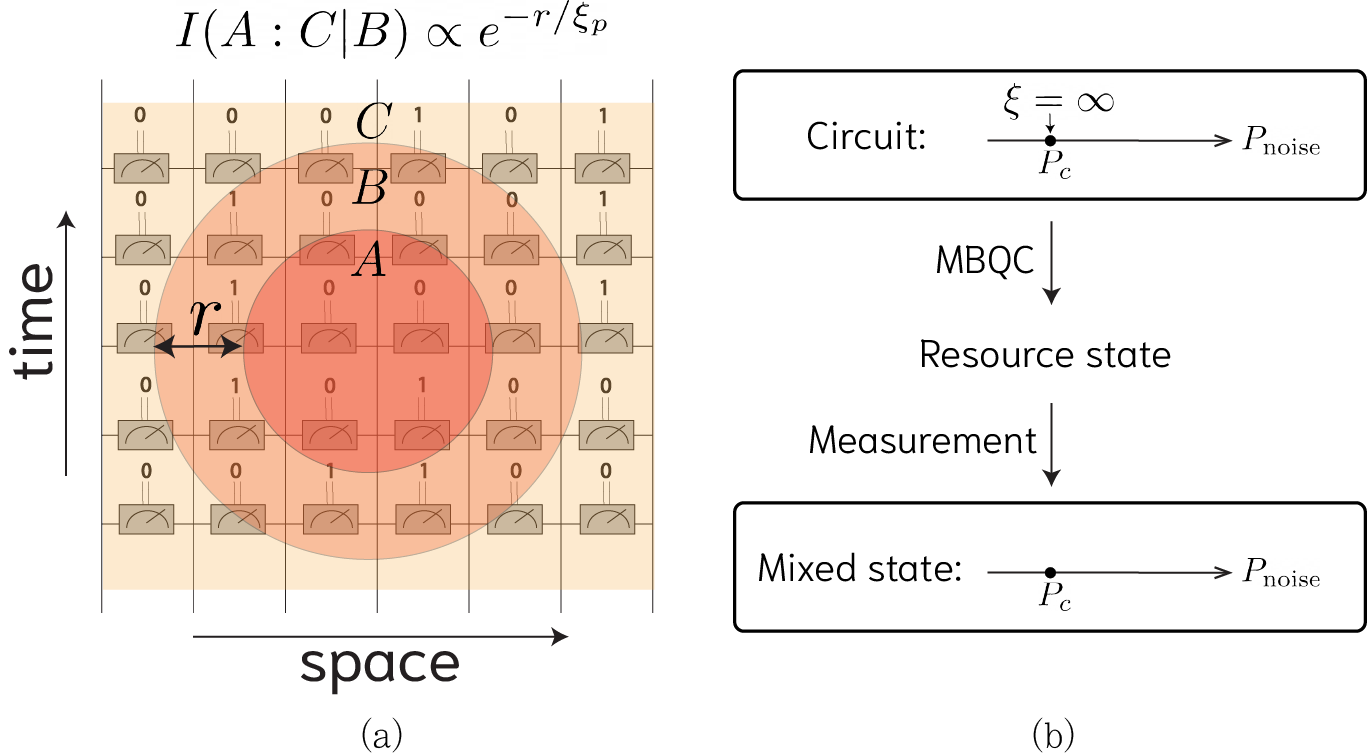}
    \caption{(a) A schematic of computing the CMI of a circuit with repeated syndrome measurements. The ensemble of measurement outcomes in spacetime is tri-partitioned, and its CMI is given by \( I(A:C|B) = H(AB) + H(BC) - H(B) - H(ABC) \), where \( H(X) \) denotes the Shannon entropy of the bits within region \( X \). It decays with a length scale called the spacetime Markov length \( \xi_p \). (b) The mapping of the error threshold to a mixed-state phase transition. The circuit can be mapped to a resource state for MBQC. Measurements of the resource state decohere the state to create a mixed state. Noise in the circuit maps to noise acting on the mixed state, which can drive a mixed-state phase transition occurring at the same point as the circuit threshold. The Markov length diverges at the mixed-state phase transition, signaling a similar divergence of the spacetime Markov length at the fault-tolerance threshold.}
    \label{fig:summary}
\end{figure}


In this work, we leverage ideas from MBQC to map stabilizer codes with both physical and measurement errors to the mixed-state phases of decohered resource states. One key insight gained from this mapping is that methods used to probe the mixed-state phase transition can be employed as diagnostics for fault tolerance. In particular, the conditional mutual information (CMI), defined as $I(A:C|B)  =S(AB) + S(BC)- S(B) - S(ABC)$ for the annular tripartition in Fig.~\ref{fig:summary}, has proven useful for detecting mixed-state phase transitions~\cite{sang2024stabilitymixedstatequantumphases}. In particular, when the CMI decays exponentially $I(A:C|B) \sim e^{-\text{dist}(A, C)/\xi}$, it defines a length scale $\xi$, 
dubbed the Markov length. The Markov length detects mixed-state phase transitions driven by local Lindbladians such as local dephasing; it is finite within each mixed-state phase but diverges at transitions~\cite{sang2024stabilitymixedstatequantumphases}. The mapping motivates an analogous quantity for the stabilizer circuit, defined by the CMI of the (classical) syndrome distribution (Fig.~\ref{fig:summary}). We show that through the mapping afforded by MBQC, the divergence of the associated length scale, which we call spacetime Markov length, signals the breakdown of fault tolerance. This correlation measure does not depend on any choice of decoder and probes the intrinsic fault-tolerance threshold.    



The organization of the text is as follows. We first provide a definition of mixed-state phases in terms of local Lindbladian evolutions and in particular, emphasize that CMI is an important measure for detecting transitions \cite{sang2024stabilitymixedstatequantumphases}. We then review the foliation of syndrome-extraction circuits into resource states, demonstrating that faults in the original circuit correspond to applying noise to the resource state. Implementing the circuit corresponds to measuring all the bulk qubits of the resource state, thus leading to a decohered resource state due to both noise and measurements. 

We then link the fault tolerance of the circuit to the ability to reverse the effects of noise on the decohered resource state.   
This allows us to identify the fault tolerance of the circuit with the condition that the decohered resource state is in the same mixed state phase as the noiseless, albeit measurement-decohered, resource state. 
Subsequently, we map the CMI of the decohered mixed state back to the circuit model, in which it becomes the CMI of the classical data extracted from the spacetime history of syndrome measurements. This sets a length scale---the spacetime Markov length---which diverges at the error threshold. Finally, we discuss the experimental relevance of this diagnostic as well as applications of decoherence in probing higher-form symmetry-protected topological (SPT) phase transitions. \\



\begin{figure*}[ht]
  \includegraphics[width=\textwidth]{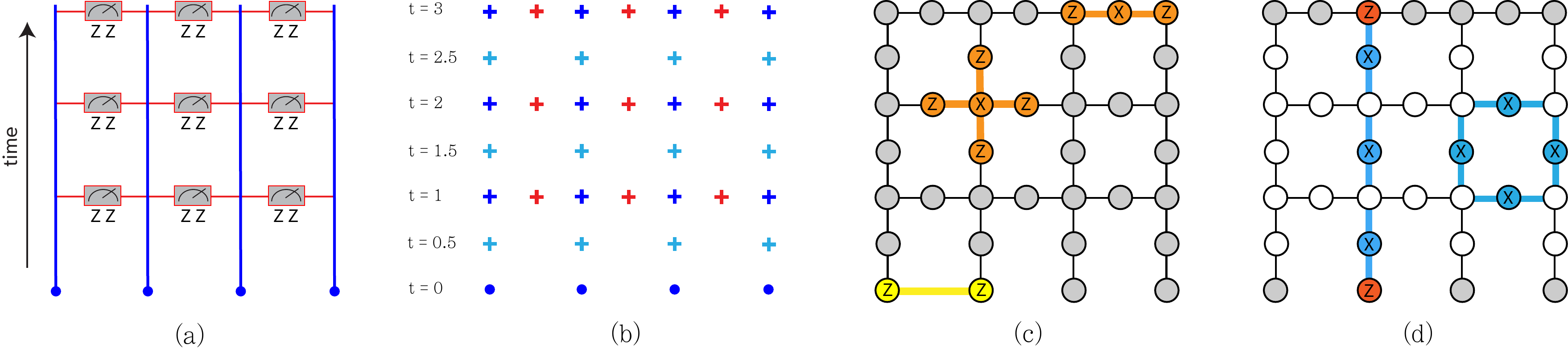}
    \caption{Mapping the syndrome-extraction circuit of the 1D repetition code to a 2D mixed state. (a) The syndrome-extraction circuit is composed of $Z_i Z_{i+1}$ measurements. (b) The resource state for MBQC is defined on copies of code qubits (blue) and syndrome qubits (red). The layers are labeled by integers and half-integers. There are no syndrome qubits in the $t=0$ layer, as the input code state is assumed to be perfect. (c) The qubits are entangled into a cluster state by applying CZ gates to neighboring qubits, with stabilizers indicated. (d) Measuring the bulk in the $X$ basis produces a mixed state coupled to the original code on the boundaries: the resulting ensemble is stabilized by local plaquette operators and composites of the logical operators of the bulk and boundary codes.}
  \label{fig:foliation}
\end{figure*}



\noindent \textbf{Mixed state phases and Markov length:} We begin by reviewing a definition of mixed-state phases and the role played by the CMI. For pure ground states of local Hamiltonians, two states are in the same phase of matter if they can be connected by finite-time evolution with a local (time-dependent) Hamiltonian. For mixed states, it is natural to generalize from a local Hamiltonian to a local Lindbladian evolution. However, since such dynamics is generally irreversible, one requires the existence of both a forward and reverse evolution with local Lindbladians. That is, there needs to exist local Lindbladians $\mathcal{L}_1$ and $\mathcal{L}_2$ such that ${\cal T} e^{\int_0^1 {\cal L}_{1,2}(t) dt}[\rho_{1,2}] = \rho_{2,1}$ to conclude that $\rho_1$ and $\rho_2$ are in the same phase~\cite{coser2019classification}.  

Given two mixed states in the same phase, finding the two-way evolutions connecting them can be challenging, and may require real-space renormalization or quantum error correction~\cite{sang2023mixed}. However, in many contexts, one is already given a one-way evolution $\rho_p \equiv {\cal T} e^{\int_0^{t_p} {\cal L}_1(t) dt}[\rho_1]$ (e.g. a noisy $\rho_1$) and the question of whether $\rho_1,\rho_p$ are in the same phase reduces to whether the noise can be undone (is there a local Lindbladian evolution from $\rho_p\rightarrow \rho_1$). To avoid having to find a reverse connection on a case by case basis, ideally one would like a criteria for the evolving state $\rho_p$ that guarantees the existence of a reverse evolution. In other words, it is desirable to find a mixed-state analogue of a gap for ground states of local Hamiltonians; as long as the gap remains nonzero upon perturbing the Hamiltonian, the corresponding family of ground states are guaranteed to be in the same phase~\mbox{\cite{bravyi2010topological, michalakis2013stability}}.   

In Ref.~\cite{sang2024stabilitymixedstatequantumphases}, it was found that the Markov length $\xi$, i.e., the length scale at which CMI decays---plays the role of the (inverse) gap for mixed states.  Given a tripartition ABC, CMI is $I(A:C|B)  =I(A:BC)-I(A:B)$, a difference of two mutual informations $I(A:B)=S(A) + S(B)-S(AB)$.  This representation shows that CMI quantifies how much correlation between $A$ and its complement is not captured by its correlations with $B$.  CMI is intimately tied to the recoverability of a state from a subsystem: when CMI is small, there exists an approximate recovery channel ${\cal R}_{B\rightarrow AB}$ such that $\rho_{ABC}\approx {\cal R}_{B\rightarrow AB}[\rho_{BC}]$.  Hence, if CMI decays rapidly, then a local channel acting on $A$ can be reversed by a channel acting on a relatively small buffer region $B$.  Concretely, Ref.~\cite{sang2024stabilitymixedstatequantumphases} used the approximate Petz theorem to argue that any local Lindbladian evolution in which the Markov length remains finite can be reversed by another local Lindbladian, where the locality of the reversal is determined by the Markov length along the evolution path.  Thus, if a state evolving under a local Lindbladian has a finite Markov length, then all states along the path remain within the same mixed-state phase.   

As an example of a mixed state phase transition, consider the two-dimensional toric code subject to bit-flip noise.  The toric code Hamiltonian is $H_{\rm t.c.}= -{\sum}_{\square} A_\square  - {\sum}_{+} B_+$, where $A_\square=\prod_{i\in\square} X_i$ and $B_+ \equiv \prod_{i\in+}Z_i$. With periodic boundary conditions, its ground state subspace is four-dimensional, encoding two logical qubits with logical operators ${\tilde X}_{v,h}, {\tilde Z}_{v,h}$ consisting of products of Pauli $X$ and $Z$ operators, respectively. Given perfect measurements of the syndromes $A_\square,B_+$ and uniform bit-flip noise with probability $p$, for $p>p_c\approx 0.11$, the logical ${\tilde Z}$ information is no longer decodable from the measurement outcomes. However, logical ${\tilde X}$ commute with the noise channel and are thus decodable for all $p$.         

This decodability transition manifests as a mixed state transition in the following sense. Given a ground state of the toric code $\ket{\rm t.c.}$, 
the bit-flip noise converts it into a mixed state $\rho_p\equiv{\calE}_p^x[\ket{\rm t.c.}\bra{\rm t.c.}]$.  Here, ${\calE}_p^x[\cdot]\equiv(1-p)(\cdot)+pX(\cdot)X$ is a bit-flip channel on every site that can be represented by a continuous Lindbladian evolution $\calL[\rho]=\sum_i\frac{1}{2}(X_i\rho X_i - \rho)$ for time $t_p=-\ln(1-2p)$.  The Markov length of the mixed state $\rho_p$ is finite everywhere except $p_c\approx 0.11$, at which it diverges \cite{sang2024stabilitymixedstatequantumphases}.  Hence, the mixed state phase transition as defined by two-way local Lindbladian connectivity to the initial state $\rho_0$ coincides with the decodability transition. 

For $p>p_c$, the fact that Markov length is finite implies that $\rho_p$ are in the same phase as $\rho_{cl}\equiv \rho_{p=0.5}=\big(\prod \frac{1+A_{\square}}{2}\big)\rho_{\tilde{X}}$, which is an ensemble of loop configurations in the $X$ basis and $\rho_{\tilde{X}}$ depends on the initial logical state.  The classical ensemble with definite logical $\tilde{X}$ operators (e.g. $\tilde{X}_{v,h}=+1$) is preparable from a product state by a local Lindbladian \cite{sang2024stabilitymixedstatequantumphases} and thus in the trivial phase under the definition in \cite{coser2019classification}. 


However, the classical ensemble $\rho_{cl}$ is non-trivial in several senses, which will be important for the purposes of our work. Most importantly, it is a topological classical memory with logical information given by ${\tilde X}_{v,h}=\pm 1$, and the logical states ($\prod \frac{1+A_{\square}}{2}\prod_{l=v,h}\frac{1\pm {\tilde X}_l}{2}$) are locally indistinguishable. This topological degeneracy is preserved under any local Lindbladian evolution in which Markov length remains finite~\cite{wip}; the approximate Petz recovery for such an evolution only depends on local reduced density matrices which are the same for all topologically degenerate states. This reverse evolution thus serves as a decoder that recovers the initial state (and any logical information encoded therein) \cite{sang2024stabilitymixedstatequantumphases}. Hence, any finite-time Lindbladian evolution connecting non-trivial to trivial topological degeneracy must produce a diverging Markov length at some point~\cite{wip}.  

This motivates the following refined definition for mixed state phases: 
\begin{definition}[\cite{wip}]\label{def: markov_path}
$\rho_{1,2}$ are in the same mixed state phase if $\rho_2 = {\cal T} e^{\int_0^1 {\cal L}(t) dt}[\rho_{1}]$ for a local Lindbladian ${\cal L}$ {\it and} ${\cal T} e^{\int_0^{t'} {\cal L}(t) dt}[\rho_{1}]$ has finite Markov length for all $t'\leq 1$.
\end{definition}
\noindent Under this refined definition, the classical state $\rho_{cl}$ is non-trivial. Indeed, upon applying a dephasing channel ${\calE}_p^z[\cdot]\equiv(1-p)(\cdot)+pZ(\cdot)Z$, the state $\rho_{cl,p}\equiv{\calE}_p^z[\rho_{cl}]$ undergoes a phase transition to a trivial phase at $p_c\approx 0.11$, at which the Markov length diverges.\\  
  


\noindent \textbf{Mapping circuits to mixed states:} 
Here, we describe how a circuit for syndrome extraction can be mapped to a mixed state. The first step is to map the circuit to a resource state for MBQC. This can be accomplished using the notion of foliation, introduced in Refs.~\cite{Bolt_2016, Brown_2020}. For simplicity, we restrict our attention to syndrome-extraction circuits for stabilizer codes and subsystem codes in $D$ dimensions with geometrically local checks. In this case, foliation produces a resource state in $D+1$ dimensions, which is short-range entangled in the bulk. Performing measurements on the resource state then reproduces the effects of the circuit and produces a mixed state according to the distribution of measurement outcomes. Note that, for simplicity, we consider Calderbank-Steane-Shor (CSS) codes throughout the text. We fully expect that the discussion can be extended to non-CSS codes using the generalizations in Ref.~\cite{Brown_2020}.

\textit{Circuit to resource state --} Following Ref.~\cite{Brown_2020}, the resource state for MBQC is defined in a Hilbert space composed of two layers of qubits for every implementation of the circuit. This is depicted in Fig.~\ref{fig:foliation} for the 1D repetition code. The two layers correspond to the $Z$-checks and $X$-checks of the syndrome-extraction circuit, respectively. We refer to the layers by integers and half-integers, so that the $m$ and $m+\frac12$ layers define the $m^{\mathrm{th}}$ application of the circuit, for $m \in \mathbb{Z}$. We take the initial implementation of the circuit to be the $t=0$ and $t=\frac12$ layers, while the final implementation is $t=m_f$, for some $m_f \in \mathbb{Z}^+$. Note that we do not include a layer for $m_f+\frac12$.\footnote{This is primarily so that, if the resource state is extended to larger $t$, we can take $m_f$ to be the initial integer layer.} 


Each layer is comprised of two types of qubits. First, each layer hosts a copy of the code qubits, so that the Hilbert space of the input code is reproduced in each layer. Second, there is a syndrome qubit for every check, with the placement of the syndrome qubit depending on the type of check. For the $Z$-type checks, we place the syndrome qubit in the integer layers, while for the $X$-type checks, the qubit is placed in the half-integer layers. Note that, for classical codes (with only $Z$-type checks), such as the repetition code, there are no syndrome qubits in the half-integer layers.

To build the resource state, we next need to specify an initial state for the code qubits in the $t=0$ layer. The choice of initial state is dependent on the context. If, for example, the circuit is being used for state preparation, then the initial state may be a product state. Here, we take the initial state to be a code state, and interpret the circuit as implementing a quantum memory. We assume that the initial code state has been prepared fault tolerantly, prior to applying the syndrome-extraction circuit.

The final step is to prepare all of the qubits (except for the code qubits in the $t=0$ layer) in the $|+\rangle$ state and couple them together with $CZ$ gates. Specifically, we apply a $CZ$ gate between each code qubit in the layer $t$ and its copies in the neighboring $t-\frac12$ and $t+\frac12$ layers.\footnote{Here, $t$ ranges from $0<t<m_f$.}
We also apply a $CZ$ gate between each syndrome qubit and the code qubits of the corresponding check. Thus, in the bulk, the state is a graph state. This completes the construction of the resource state. The resource state for the 1D repetition code is shown in Fig.~\ref{fig:foliation}.

\textit{Resource state to mixed state --} Now that we have defined the resource state for MBQC, we are able to finish the mapping of the syndrome-extraction circuit to a mixed state. In practice, the syndrome-extraction circuit is implemented within the MBQC formalism by measuring all of the qubits in the $X$ basis, except for the code qubits in the final layer, that is, $t=m_f$. Given that the measurement outcomes are probabilistic, this produces a mixed state that is a classical distribution of the $X$ basis states (except on the code qubits at $t=m_f$). 

The effect of measuring the single-site $X$ operators is equivalent to adding full-strength bit-flip noise to the resource state.\footnote{This can be seen directly for a single qubit in the state $\rho$, since the post-measurement state is 
$\frac{1+X}{2}\rho\frac{1+X}{2} + \frac{1-X}{2}\rho\frac{1-X}{2}$, which is equal to $\frac12 \rho + \frac12 X\rho X$.} Therefore, letting $\rho_{\mathrm{RS}}$ denote the resource state, the mixed state obtained after the measurements is:
\begin{align}
    \rho_{0}\equiv\mathcal{E}^x_{p=\frac{1}{2}} (\rho_{RS}).
\end{align}
Here, the channel $\mathcal{E}^x_{p=\frac{1}{2}}$ is the full-strength bit-flip channel, which, for each site, applies a Pauli $X$ operator with probability $p=\frac12$. Implicitly, $\mathcal{E}^x_{p=\frac{1}{2}}$ acts on every qubit except for the code qubits in the final layer $t=m_f$.

To gain intuition for the mixed state $\rho_0$, we consider the bulk state $\rho_{\mathrm{cl}} = \mathrm{Tr}_{m_f}(\rho_0)$, where we have traced out the top layer $t=m_f$. The state $\rho_{\mathrm{cl}}$ is a purely classical state. This is because, due to the full-strength bit-flip noise, the state $\rho_{\mathrm{cl}}$ commutes with every single-site $X$ operator. Therefore, $\rho_{\mathrm{cl}}$ is diagonal in the $X$ basis, i.e., it is a mixture of $X$ basis states. For example, starting with the 1D repetition code, the state $\rho_{\mathrm{cl}}$ is a mixture of loop configurations, where the loops are formed by $|-\rangle$ states on the dual lattice. We argue below that the threshold of the underlying quantum code coincides with a phase transition in the bulk of the mixed state $\rho_0$. \\

\noindent \textbf{Mapping noisy circuits to noisy states:} 
We demonstrate that noise in the circuit model corresponds to the evolution of the mixed state $\rho_0$ under a particular local Lindbladian. This naturally motivates the question of whether the mixed state undergoes a phase transition induced by the Lindbladian evolution. In the sections that follow, we argue that the question is equivalent to identifying whether a threshold exists for the syndrome-extraction circuit and furthermore that the mixed-state phase transition coincides with the threshold.

Consider an \( X \) or \( Z \) error occurring on a code qubit \( i \) between two consecutive implementations of the circuit, denoted by \( t \) and \( t+1 \). As discussed previously, each time step \( t \) in the circuit corresponds to two time steps in the resource state, specifically \( t \) and \( t + \frac{1}{2} \). This mapping yields the following translation of errors:

\begin{enumerate}
    \item \textbf{Circuit \( X \) error:} This becomes a \( Z \) error in the resource state at time step \( t + \frac{1}{2} \) on the corresponding code qubit \( i \).
    
    \item \textbf{Circuit \( Z \) error:} This becomes a \( Z \) error in the resource state at time step \( t + 1 \) on the corresponding code qubit \( i \).
    
    \item \textbf{Readout error:}  This becomes a \( Z \) error on the corresponding syndrome qubit of the resource state.
\end{enumerate}


\begin{figure}[h]
\includegraphics[width=0.48\textwidth]{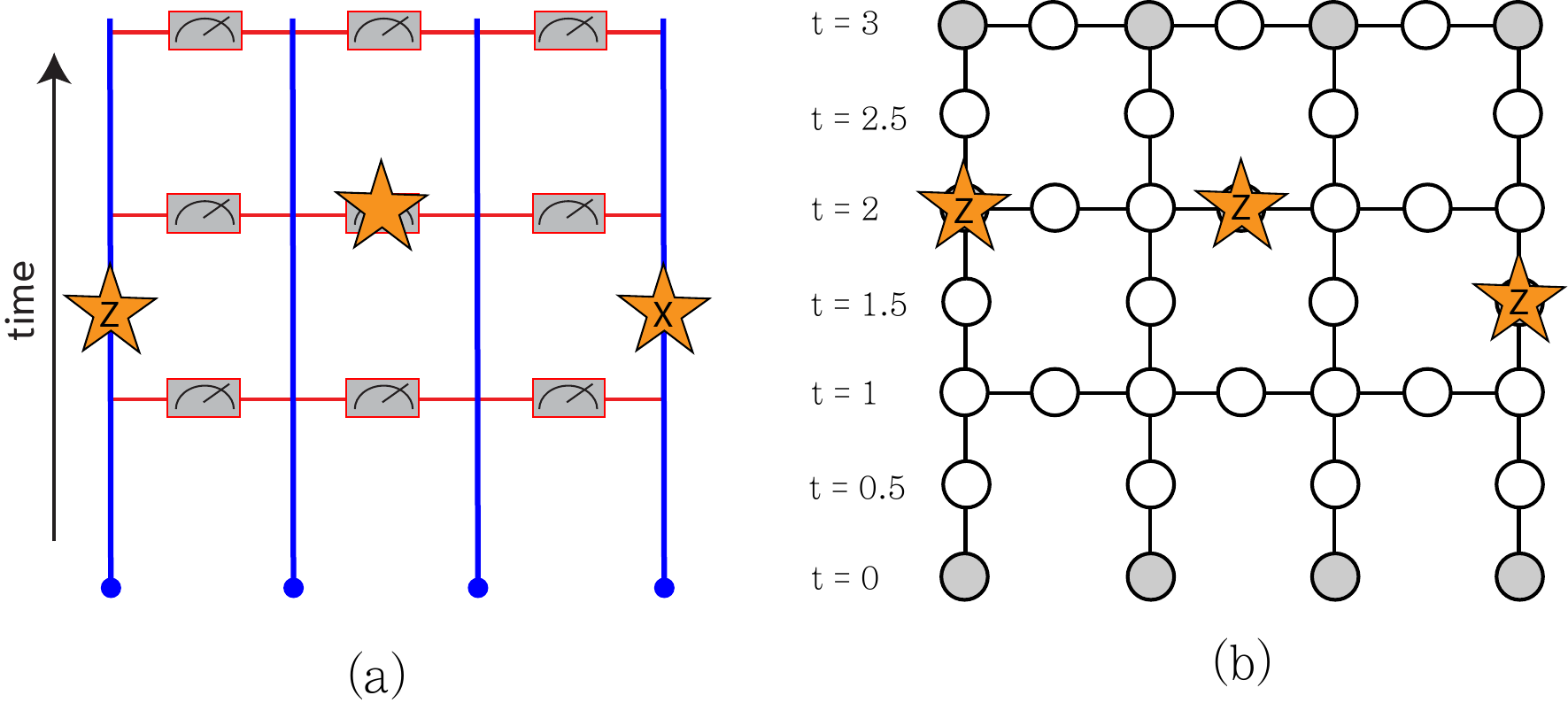}
    \caption{Error mapping from the syndrome-extraction circuit to the resource state for the 1D repetition code. Error locations are denoted by stars and are translated to the MBQC setting according to rules 1.-3.~in the main text.  
    }
    \label{fig:Error}
\end{figure}

To understand this noise mapping, it is instructive to examine how qubit \( i \) is teleported from time \( t \) to \( t+1 \). 
As described in Ref.~\cite{Brown_2020, hauser2024informationdynamicsdecoheredquantum}, this is achieved by measuring this code qubit at times \( t \) and \( t+\frac{1}{2} \). After these two measurements, the state of code qubit \( i \) at time \( t \) is teleported to time step \( t+1 \), though with a Pauli correction depending on the measurement results. Let us denote the two measurement results by \( s_{i,t} \) and \( s_{i,t+\frac{1}{2}} \), where they take the value 0 when the outcome is the $|+\rangle$ state and 1 when the outcome is the $|-\rangle$ state. These measurement results alter the teleported state at time \( t+1 \) by the action of the Pauli byproduct operators:
\begin{align} \label{eq: byproduct operators}
Z^{s_{i,t}} X^{s_{i,t+\frac{1}{2}}}.
\end{align}

This demonstrates that changing the value of the measurement result at integer time will manifest as an additional \( Z \) operator acting on the teleported state, while changing the result at half-integer time will result in an additional \( X \) operator acting on the state. The change in the measurement results can be viewed as a \( Z \) error on a code qubit, as this flips the value of $s$. Furthermore, a \( Z \) error on the syndrome ancilla flips the outcome of the \( X \) measurement, manifesting as a readout error. This unifies all stochastic circuit errors as correlated \( Z \)-dephasing errors in the resource state. 


If these errors occur randomly in the circuit model with probability \( p \), they will similarly manifest in the resource state with probability \( p \). (For simplicity, we assume Pauli noise and readout error occur with the same probability, but the mapping readily generalizes.) Therefore, the resource state undergoes a dephasing channel 
\(
\mathcal{E}^z_{p}(\cdot) = (1 - p) (\cdot) + p Z (\cdot) Z,
\)
on the syndrome qubits and code qubits depending via the above map on the type of Pauli error.  The resulting mixed state after both noise and measurement is
\begin{align}
\rho_p \equiv \mathcal{E}^x_{p=\frac{1}{2}} (\mathcal{E}^z_{p}(\rho_{RS})) = \mathcal{E}^z_{p}(\rho_0),
\label{eq:one way connection}
\end{align}
as $\mathcal{E}^z_{p},\mathcal{E}^x_{p=1/2}$ commute. \\

\noindent \textbf{Fault tolerance and mixed-state phase transition:} \\
We now establish a connection between the fault tolerance of the syndrome-extraction circuit and the mixed-state phases of the measured resource state. 
We begin by showing that the fault tolerance of the circuit is equivalent to the fault tolerance of the corresponding MBQC protocol. 



We consider a circuit or an MBQC protocol fault tolerant if they admit an error threshold for successful recovery (with a realistic error model) and they are implemented below threshold for a sufficiently large system size. 
Given a noisy circuit with error rate \(p\), we let $\tilde{\rho}_p = \sum_{\vec{S}} p_{\vec{S}}  |\vec{S}\rangle\langle\vec{S}| \otimes \rho_{t,\vec{S}}$ denote the mixture of all possible syndromes $\vec{S}$, weighted by their probabilities $p_{\vec{S}}$, with the corresponding final states $\rho_{t,\vec{S}}$ of the circuit.
Fault tolerance then requires the existence of a recovery map \( \tilde{R} \) such that $\tilde{R}(\tilde{\rho}_p) \approx \tilde{\rho}_0$ for $p < p_c$, 
where \( p_c \) denotes the noise threshold and \( \tilde{R} \) is independent of the encoded information.\footnote{This recovery condition assumes that the fault-tolerant circuit correct all errors. In practice, there can be residual uncorrected errors that can be handled in future iterations of the circuit. Hence we disregard these residual errors by assuming that no errors occur in the final round of the circuit. Equivalently, for MBQC, this corresponds to the noise channel \( \mathcal{E}^z_{p} \) acting on all qubits except those at the top boundary.}  Similarly, fault tolerance in the MBQC setting requires the existence of  a recovery map \(R\) such that
$R(\rho_p) \approx \rho_0$ for $p < p_c$.


Fault tolerance in the circuit and MBQC setting are equivalent because the existence of a recovery channel \(R\) yields a recovery channel $\tilde{R}$ for the corresponding circuit, and vice-versa. To see this, we first note that the state $\tilde{\rho}_p$ corresponds to the state $\rho_p$ when the code qubits are post-selected in the \(|+\rangle\) state within the bulk. This post-selection can be implemented through a feedback channel \(F\), such that \(F(\rho) = \tilde{\rho}\). The feedback channel measures all bulk code qubits, obtaining outcomes $\vec{C}$, and applies unitary corrections $U_{\vec{C}}$ to all syndrome and code qubits, conditioned on the outcomes. 
The feedback channel can be inverted \((\rho = F^{-1}(\tilde{\rho}))\), where \(F^{-1}\) 
applies  $U_{\vec{C}}^{-1}$ with probability \(1/2\).
Therefore, the circuit and MBQC recovery maps are related by:
\begin{align}
\tilde{R} = F \circ R \circ F^{-1}.
\label{eq: recovery}
\end{align}
For further details on the feedback channel $F$, we refer to Appendix~\ref{app:A}.
 

This allows us to relate the threshold of the circuit to the mixed-state phase transition of $\rho_p = \mathcal{E}^z_{p}(\rho_0)$. These states are always one-way connected from $\rho_0$ by the noise channel $\mathcal{E}^z_{p}$, and the question of whether they are in the same phase thus hinges on the existence of a recovery $R$, which guarantees a circuit recovery $\tilde{R}$ as we showed. 

We first prove by contradiction that the Markov length diverges at the error threshold for the circuit. Assume instead that the Markov length remains finite at \( p_c \) and within an \( \epsilon \)-neighborhood around $p_c$. Given the noise channel evolving the density matrix $\rho_p$ from \( p = p_c - \epsilon \) to \( p = p_c + \epsilon \), a finite Markov length in this region implies that a reverse Lindbladian can be constructed, evolving the density matrix from \( p_c + \epsilon \) back to \( p_c - \epsilon \). This reverse Lindbladian evolution, when combined with the recovery map $R$ below the threshold, would enable recovery\footnote{Note that the combination of the reverse Lindbladian and a sub-threshold recovery is a valid recovery because the approximate Petz recovery of the reverse Lindbladian depends solely on local reduced density matrices, which are independent of the encoded logical information of the underlying topological code.} above \( p_c \), contradicting the assumption that \( p_c \) represents the true threshold.

While this argument is sufficient to show that the Markov length diverges at \( p_c \), it does not exclude the possibility of divergence at multiple points. However, we expect that this generically does not happen, based on mapping the problem of recovery to statistical-mechanics models \cite{dennis2002topological}. In this mapping, the (dis)ordered phase of the stat mech model corresponds to the (un)recoverable regime. The CMI corresponds to the free energy cost of a point defect in the stat mech model~\cite{sang2024stabilitymixedstatequantumphases}, and thus, the Markov length can be identified with the correlation length---assuming that there is only one diverging length scale in the vicinity of the critical point of the stat mech model. As a consequence, the divergence of the Markov length coincides with the recoverability transition. 

For simplicity, we have considered incoherent noise on the syndrome extraction circuit. In the case of non-stochastic errors, such as coherent errors, the mapping of the errors to the resource state is more subtle. In Appendix~\ref{app:B}, we discuss the subtleties and show that for the repetition code in particular, coherent noise on physical qubits in the circuit model is mapped to incoherent noise on the resource state -- allowing for a connection to a mixed-state phase transition. We also note in Appendix~\ref{app:C} that coherent noise on the MBQC resource state is equivalent to $Z$-dephasing errors on the $X$-dephased resource state. This implies that the threshold of MBQC to coherent noise corresponds to a mixed-state phase transition of the $X$-dephased resource state under incoherent noise.\\

\noindent \textbf{Classical memory in the bulk:}  Here we analyze the above mixed-state phase transition in more detail and 
connect the fault tolerance of the circuit to the recoverability of a classical memory in the bulk of the decohered resource state. 
We identify two important classes of stabilizers of the resource state, which we refer to as \emph{detector cells} and \( L'BL \), which define the bulk classical memory. 

\textit{Detector cells --} The first type of stabilizers, referred to as {detector cells} following Refs.~\cite{Kesselring2024condensation, Paesani2023foliation}, capture the redundancy inherent to the circuit. Without errors, consecutive measurements of a stabilizer yield consistent syndromes. For instance, in the repetition code, the measurements of the stabilizer \( Z_i Z_{i+1} \) at times \( t \) and $t+1$ should agree in the absence of errors. In MBQC, this process is equivalent to measuring the \( X \) operator on the syndrome qubit associated with the stabilizer. Measuring \( X_i \) and \( X_{i+1} \) on the code qubits in the intermediate layer (\( t + \frac{1}{2} \)) teleports the stabilizer from \( t \) to \( t+1 \), possibly with a sign change $X_i X_{i+1}$. Finally, measuring \( X \) on the syndrome qubit at \( t+1 \) yields a second syndrome measurement, influenced solely by the teleportation-induced sign change. Thus, assuming no errors, the product of these \( X \) measurements should yield \( +1 \). This detector cell is depicted in Fig.~\ref{fig:foliation}(d) as a plaquette of \( X \) operators.

In general, for a stabilizer \( S \) measured in the circuit, the corresponding detector cell is:
\begin{eqs} \label{eq: detector cell}
    &D_S = \\ &\left( \prod_{i \in \mathrm{Ch}(S)}  X_{i}^{(t+1)} \right) \left(  \prod_{i \in \mathrm{Sup}(S)} X_i^{(t+\frac12)} \right) \left( \prod_{i \in \mathrm{Ch}(S)}  X_{i}^{(t)} \right).
\end{eqs}
Here, $\mathrm{Ch}(S)$ is a choice of syndrome qubits such that the product of the corresponding checks is $S$,\footnote{Note that the choice of syndrome qubits is for subsystem codes in particular, where different choices of gauge operators may multiply to the same stabilizer.} and \( \mathrm{Sup}(S) \) is the set of code qubits in the support of \( S \). Measuring \( X \)-operators on \( \mathrm{Sup}(S) \) in the intermediate layer \( t+\frac{1}{2} \) teleports the stabilizer \( S \) from layer \( t \) to \( t+1 \), with the \( X \)-measurements at \( t+1 \) providing a second syndrome consistent with that at \( t \).\footnote{Detector cells can also correspond to local redundancy checks within a single layer. For example, in the 3D toric code, there are detector cells represented by the product of plaquette stabilizers around a cube. However, we do not explicitly discuss these types of detector cells here, for simplicity.}

\textit{\( L'BL \) stabilizers --} The second type of stabilizer arises from the temporal consistency of logical operators. Without noise, the logical information encoded in the code at different time instances should agree. Consider two instances of a repetition code logical operator, \( Z^{(m_f)}_i \) and \( Z^{(0)}_i \), acting on the code qubits in the final and initial layers of the circuit. The temporal consistency of this logical is reflected in the resource state through a stabilizer of the following form:
\[
    L^Z_{m_f} B_{L^Z} L^Z_{0} = Z^{(m_f)}_i \left( \prod_{k=0}^{m_f-1} X_i^{(k+\frac{1}{2})} \right) Z^{(0)}_i,
\]
Here, the logical operators \( L^Z_{m_f} \) and \( L^Z_{0} \) are linked via the operator \( B_{L^Z} \), which acts within the bulk of the state. This stabilizer is illustrated in Fig.~\ref{fig:foliation}(d), where the supports of \( L^Z_{m_f} \) and \( L^Z_{0} \) are highlighted in red, and the support of \( B_{L^Z} \) is depicted in blue. 

In general, the resource state has stabilizers of the form \( L_{m_f} B_{L} L_{0} \), with \( B_L \) given by
\begin{eqs} \label{eq: classical logical}
     B_{L^Z} &= \prod_{k=0}^{m_f-1}\prod_{i \in \mathrm{Sup}(L^Z)}X_i^{(k+\frac12)}, \\
     B_{L^X} &= \prod_{k=0}^{m_f-1}\prod_{i \in \mathrm{Sup}(L^X)}X_i^{(k)},
\end{eqs}
depending on whether the logicals $L_0, L_{m_f}$ are products of Pauli $Z$ or $X$ operators. We refer to the operators of the form \( L_{m_f} B_{L} L_{0} \) as $L'BL$ stabilizers, for simplicity.

\textit{Classical memory in the bulk --} We now explain how the \( B_L \) and \( D_S \) operators together define a classical memory. In this memory, the \( D_S \) operators act as the check operators, while the \( B_L \) operators represent the logical bits. Both \( D_S \) and \( B_L \) are products of \( X \)-type operators and thus commute. Moreover, the \( B_L \) operators are independent of the \( D_S \) operators because logical bits represented by \( B_L \) can be flipped by conjugating them with \( L_m \) or \( L_{m-\frac{1}{2}} \),\footnote{Here, $L_m$ and $L_{m-\frac{1}{2}}$ are the logical operators of the underlying code defined within the $m$ and $m-\frac12$ layers of the resource state, respectively.} while the \( D_S \) operators remain invariant (a logical error does not alter the values of detector cells).

For example, the bulk code derived from the repetition code has plaquettes of \( X \)-type operators serving as checks and open strings of \( X \)-type operators connecting one boundary to another as logical operators, as depicted in Fig.~\ref{fig:foliation}. The code distance scales linearly with the system size, as non-contractible closed loops of \( Z \)-errors can alter the logical bit without being detected by the checks. Note that for the 1D repetition code, there is only one bit encoded, corresponding to the operator \( B_{L^Z} \). In contrast, there is no bit corresponding to the operator \( B_{L^X} \), as the repetition code is designed to protect against \( X \)-type noise only.

\textit{Error correction and fault tolerance --} We now analyze how the error correction properties of this classical code relate to the fault tolerance of the circuit. Let us begin with the noiseless scenario. In this case, transmitting a logical state from one boundary to another in MBQC can be understood as a two-step process. First, measurements in the bulk establish entanglement between the top and bottom boundaries. Then, measuring the bottom boundary teleports the information to the top boundary, upon applying feedback depending on all measurement outcomes.

The \( L'BL \) stabilizer enables this teleportation process. The initial state satisfies \( L'BL = 1 \), and measuring the bulk corresponds to measuring the \( B \) operators. Consequently, the state on the boundaries is stabilized by \( L'L \), taking value $B$. Successfully teleporting the logical state from one boundary to the other requires feedback depending on the measured value of \( B \).

In the presence of noise, the errors need to be removed while preserving \( L'BL \); otherwise the feedback is wrong.  The condition for removing errors without changing a logical operator $B$ is precisely the condition for the classical memory to persist in the presence of noise. 
Thus, retention of classical memory implies the ability to teleport the logical state from the initial boundary to the final boundary, which in turn implies preservation of logical information in the original circuit.

As an example, the threshold for the 1D repetition code with probability $p$ for both bit-flip and readout error maps to the threshold for the 2D classical memory defined by $\rho_{cl}=\big(\prod \frac{1+A_{\square}}{2}\big)\rho_{\tilde{X}}$ subject to dephasing noise ${\calE}_p^z$.  We note that \cite{zhang2024quantumcommunicationmixedstateorder} offers a complementary approach to our mixed state phase perspective, using strange correlators of the noisy cluster state to diagnose its ability to transmit information from one boundary to the other.

\textit{Higher-Form Symmetries --} In some cases, the detector cells generate higher-form symmetries, i.e., for which the symmetry operators are supported on closed submanifolds, as described in Ref.~\cite{Qi2021higherform}. In particular, when the stabilizers generate a \( q \)-form symmetry, the detector cells then generate a \( q \)-form symmetry in one dimension higher. For example, the stabilizers of the toric code generate a \(\mathbb{Z}_2 \times \mathbb{Z}_2\) 1-form symmetry. Correspondingly, the detector cells of the foliated state also generate a \(\mathbb{Z}_2 \times \mathbb{Z}_2\) 1-form symmetry---this is the well-known \(\mathbb{Z}_2 \times \mathbb{Z}_2\) 1-form symmetry of the 3D cluster state~\cite{Roberts20201form}. Thus, in this case, the redundancy of the stabilizer measurements produces a higher-form symmetry of the resource state.

This contrasts with cases such as the Bacon-Shor code, which is not fault tolerant. In that case, the detector cells of the resource state are rigid, in the sense that they are linear subsystem symmetries of the resource state. We postulate that, if the checks are local, then higher-form symmetries of the resource state are necessary for fault tolerance.

In the case that the stabilizers of the underlying code generate an anomalous \( q \)-form symmetry, then the stabilizers of the form \( L'BL \) are an indication that the resource state belongs to a nontrivial \( q \)-form SPT phase. The \( L'BL \) operators imply that the symmetry defects of the \( q \)-form symmetry (created by \( B \)) are dressed with \( q \)-form symmetry charges, given by \( L' \) and \( L \). This is because the operator \( L_m \) (\( L_{m-\frac12} \)) defined by the logical operator \( L^Z \) (\( L^X \)) fails to commute with \( B_{L^X} \) (\( B_{L^Z} \)) when they have overlapping support and \( L^Z \) and \( L^X \) are conjugate logical operators. 

For example, the stabilizers of the toric code generate an anomalous \(\mathbb{Z}_2 \times \mathbb{Z}_2\) 1-form symmetry. The resource state is known to belong to a nontrivial 1-form SPT phase, as seen by the \( L'BL \) stabilizers. We refer to Refs.~\cite{Roberts20201form, okuda2024anomalyinflowcssfractonic} for further details.\\

\noindent \textbf{Spacetime Markov length:} We employ the Markov-length as a diagnostic for a mixed-state phase transition to show that the CMI of the classical data extracted from a noisy circuit serves as an indicator of the error threshold. We begin by showing that the CMI of a decohered resource state depends exclusively on the entropy of specific detector cells. We then demonstrate a parallel result for the circuit itself, where the CMI of classical syndromes depends solely on the entropy of the corresponding classical bits associated with the detector cells. Finally, we prove that under the noise mapping, the entropy of the resource state’s detector cells matches the entropy of their circuit counterparts. This correspondence establishes an equality between the CMI of the decohered resource state and that of the classical data, thereby confirming spacetime Markov length as a robust indicator of the error threshold. 


We begin by examining a decohered resource state derived from the correspondence with a given quantum circuit. To quantify the CMI of this state, we compute the reduced density matrix for a given subregion $A$  in the bulk of the decohered resource state and analyze its entropy. Decohering the system in the \( X \)-measurement basis effectively removes all initial stabilizers of the state except the ones that commute with \( X \)-measurements. These stabilizers are composed of Pauli-\( X \) operators and are the detector cells.
Consequently, the reduced density matrix post-measurement (with discarded outcomes) assumes the form:
\begin{equation}
    \rho_A = \frac{1}{N} \prod_k \frac{1 + D_k}{2},
\end{equation}
where \( N \) is a normalization factor, and \(D_k \) are stabilizers corresponding to detector cells inside the region.\footnote{In general, non-local symmetries may exist within a given region, particularly for regions that are not simply connected; however, we still refer to these as detector cells.}

Introducing a noise channel with strength $p$ to this reduced density matrix modifies it to:
\begin{equation}
    \rho_{A,p} = \frac{1}{N} \sum_{m_k} P_{m_k} \prod_k \frac{1 + m_k D_k}{2},
\end{equation}
where \( P_{m_k} \) represents the probability distribution over the sign of detector cells (\(m_k\)) in the resource state. Computing the von Neumann entropy of this density matrix gives:
\begin{equation}
S(\rho_{A,p}) = H(m) + |A| - |D|
\end{equation}

This has two contributions: the first is the Shannon entropy of the probability distribution of the sign distribution, \( H(m) = \sum P_{m_k} \log P_{m_k} \), and the second is a term from the normalization factor \( N \), which depends on the number of qubits inside the region ($|A|$) and the number of detector cells within it ($|D|$).

Next, we examine the corresponding region \( A \) in the circuit. In a noise-free scenario, the extracted syndromes have deterministic components that correspond to detector cells in the resource state. For instance, consider two syndromes \( s_1 \) and \( s_2 \) that are outcomes of repeated measurements of a given stabilizer. From the consistency of the result in the noise-free scenario, we know \(s_1 = s_2\) which means \( d_1 = s_1 s_2\) is a deterministic variable.
Given a bit string \( s_i \) from region \( A \), the syndromes can be expressed in terms of deterministic components \( d_i \) and a basis for non-deterministic components \( n_i \). The non-deterministic syndromes remain completely random, even under stochastic errors, as seen from the stabilizer formalism. Thus, the entropy of \( s_i \) includes two contributions: one from the deterministic components \( d_i \) and another from the fully random components \( n_i \). Stochastic errors turn the deterministic components \( d_i \) into random variables, with probability distribution \( P(d_1, d_2, \ldots, d_k) \). This yields the total entropy:
\begin{equation}
    H(s) = H(d) + |s_i| - | d_i|,
\end{equation}
where \( H(d) \) represents the entropy of the \( d_i \) values, and the second two terms account for the \(|n_i|\) fully random variables within the region.

The correspondence between detector cells in the resource state and deterministic components in the circuit implies \(\ |d_i| = |D|\), and the noise mapping ensures that any error changing the sign of a detector cell also changes the sign of the corresponding deterministic component. This shows that the probability distribution of the sign of detector cells \(P_m\) matches that of the deterministic components \(P_d\), so \(H(m) = H(d)\). Because the volume law terms \( |s_i|\) and \( |A| \) do not contribute to CMI, this establishes the equality of CMI between the circuit and MBQC. \\

\noindent \textbf{Discussion:} We have shown that the breakdown of fault tolerance in local stabilizer quantum memories can be mapped to mixed state transitions of classical memories in one higher dimension. This mixed-state perspective offers the novel diagnostic of the spacetime Markov length, which can be computed directly from the syndrome distribution of the circuit. This diagnostic does not depend on any decoder and thus probes the intrinsic threshold transition. 

As repetition and surface codes below the noise threshold have been realized recently on quantum hardware~\cite{acharya2024quantumerrorcorrectionsurface}, it would be very interesting to extract the spacetime Markov length from experimental syndrome data as a complementary probe of the proximity to the threshold. Real-time decoders acting in parallel on space-time blocks (and then finding a globally consistent solution) have been demonstrated~\cite{acharya2024quantumerrorcorrectionsurface}. The spacetime Markov length scale, in particular, may be useful to estimate the minimum block sizes for such decoders.   

This approach can also be applied to more general circuits that implement logical gates through code deformations, such as lattice surgery. These circuits define resource states where the logical gates are encoded into the geometry of the cluster state~\cite{Brown_2020}. Further research could also investigate mixed-state phase transitions within the formalism of $ZX$ calculus~\cite{Bombin2024unifyingflavorsof}, which would generalize our analysis to inherently dynamical quantum-error correcting codes, such as Floquet codes~\cite{Hastings2021dynamically}. 
Another promising direction is the exploration of spacetime circuit codes, where we anticipate that the outcome code \cite{delfosse2023spacetimecodescliffordcircuits} could serve a similar role as the classical memory in our analysis. In this context, error correction for the classical memory would be linked to fault tolerance, and the spacetime Markov length could once again be used to detect the threshold. 


While this work focused on fault-tolerant stabilizer codes (and corresponding Clifford operations),
it would be interesting to extend this approach to non-Abelian topological memories or operations beyond the stabilizer paradigm. The general case requires adaptive feedback from syndrome measurement, meaning that the mapping of a fault-tolerant circuit to a noisy resource state would be more complex. However, in this context, the spacetime Markov length may still be relevant. Fault tolerance with measurement errors relies on the redundancy of the syndromes, and thus it is plausible that the classical syndrome state must generally be non-trivial and spacetime Markov length may detect its transition to a trivial state.

{\it Application to higher-form SPT transitions:} We note that the measurements which decohere the resource state into a classical ensemble are essential for defining a mixed state phase transition.  Without measurements, the resource state with periodic boundary conditions does {\it not}, in fact, undergo a phase transition when noise is applied. This is because a cluster state subject to local dephasing noise is equivalent to a thermal Gibbs state of the cluster-state Hamiltonian~\cite{PhysRevA.71.062313,chen2023symmetryenforced}, which has zero CMI~\cite{brown2012quantummarkovnetworkscommuting}. 

The measurement decoherence which exposes the mixed state phase transition may be more generally useful in exposing ``hidden'' transitions of SPT phases protected by higher-form symmetries under both incoherent and coherent perturbations. For example, in Ref.~\cite{verresen2024higgscondensatessymmetryprotectedtopological}, there are instances in which the higher-form symmetry protecting an SPT is explicitly broken, leading to no bulk phase transition but which still exhibits a boundary transition. The effect of the measurement decoherence in the symmetry-charge basis is to effectively introduce a boundary after spacetime rotation~\cite{bao2023mixedstatetopologicalordererrorfield}, thus exposing the hidden transition.    
\\

\noindent \textbf{Acknowledgments:} AN thanks Ali Lavasani for valuable discussions. TDE acknowledges Lawrence Cohen, Dongjin Lee, Yaodong Li, Sam Roberts, Charles Stahl, and Dominic Williamson for work on related projects. TH thanks Tarun Grover, Leonardo Lessa, Roger Mong, Shengqi Sang, and Chong Wang for collaboration on related projects.  This work was supported by the Perimeter Institute for Theoretical Physics (PI), the Natural Sciences and Engineering Research Council of Canada (NSERC), and an Ontario Early Researcher Award. Research at PI is supported in part by the Government of Canada through the Department of Innovation, Science, and Economic Development, and by the Province of Ontario through the Ministry of Colleges and Universities.

\bibliography{references}

\onecolumngrid

\appendix

\section{Feedback Channel}
\label{app:A}

Here, we describe the feedback channel $F$ used in the main text. The feedback channel is an important step in performing MBQC, since the measurement outcomes of the code qubits are random. If the measurements of the code qubits are postselected to be in the $\ket{+}$ state, then the state of the code qubits is seamlessly teleported to the final layer. Otherwise, if one or more code qubits are measured in the $\ket{-}$ state, the realized circuit deviates from the intended one by byproduct operators [see Eq.~\eqref{eq: byproduct operators}], which must be tracked and corrected via the feedback channel $F$. We argue here that the feedback channel allows us to map the measured resource state to the circuit model and a map $F^{-1}$ takes us back from the circuit model to the measured resource state. We start by considering the noiseless case.

\subsection{Feedback Channel for Noiseless MBQC}


We begin by introducing $F$ in the absence of any noise. Using the notation in the main text, we define the density matrix of the circuit in the noiseless limit as $\tilde{\rho}_0$. Given an initial code state $\rho_i$, measuring a syndrome and recording its outcome is described by the channel $M$, acting as follows: 
\begin{equation}
M(\rho_i) = \sum_s P_s \rho_i P_s \otimes |s\rangle\langle s|,
\end{equation}
where $P_s$ is the projection operator associated with syndrome measurement outcome $s$, and $|s\rangle\langle s|$ denotes the syndrome state. Extending this process over multiple instances of syndrome extraction, the final density matrix $\tilde{\rho}_0$ is: 
\begin{equation}
\tilde{\rho}_0 = \sum_{\vec{S}} P_{\vec{S}} \rho_i P_{\vec{S}}\otimes |\vec{S}\rangle\langle\vec{S}|,
\end{equation}
where $\vec{S}$ represents the sequence of all syndrome outcomes, and $P_{\vec{S}}$ is the corresponding projection operator sequence i.e., the product of projectors $P_{\vec{S}} = P_{\vec{S}_{t_f}} \cdots P_{\vec{S}_1}$. The state $\tilde{\rho}_0$ both encodes the final state of the circuit as well as all syndrome qubits through the history. Note that in the absence of errors, for a stabilizer code, as opposed to a subsystem code, all of the syndrome outcomes are $0$.


In contrast, the post-measurement density matrix of the MBQC resource state can be expressed as:
\begin{equation}
\rho_0 = \frac{1}{2^{|\vec{C}|}} \sum_{\vec{C}} |\vec{C}\rangle\langle\vec{C}| \otimes \tilde{\rho}_{0,\vec{C}},
\end{equation}
where $|\vec{C}\rangle\langle\vec{C}|$ corresponds to the outcomes of measuring code qubits, each equally likely. The conditional state $\tilde{\rho}_{0,\vec{C}}$ encodes both syndrome measurement outcomes and the final boundary state of the code qubits. Note that, in this case, the syndrome outcomes and state on the final layer depend on the measurement outcomes of the code qubits because of the Pauli byproducts associated with $\ket{-}$ measurements of the code qubits. Thus, $\tilde{\rho}_{0,\vec{C}}$ can be written as:
\begin{align}
\tilde{\rho}_{0,\vec{C}} = \sum_{\vec{S}} &P_{\vec{S}_{t_f}} \bigg(\dots U_{\vec{C}_3} \Big( P_{\vec{S}_2} \big(U_{\vec{C}_2}\,(P_{\vec{S}_1}\,(U_{\vec{C}_1}\rho_i U_{\vec{C}_1}^\dagger)P_{\vec{S}_1})U_{\vec{C}_2}^\dagger\big)P_{\vec{S}_2}\Big) U_{\vec{C}_3}^\dagger\dots\bigg)P_{\vec{S}_{t_f}} \otimes |\vec{S}\rangle\langle\vec{S}|,
\label{eq: adap}
\end{align}
where $U_{\vec{C}_i}$ denotes the Pauli byproducts at a given time step $i$ determined by the code qubit measurements, as written in Eq.~\eqref{eq: byproduct operators}, and $P_{\vec{S}_j}$ are projectors associated with syndrome measurements at each time step. From this expression, we now construct the feedback channel, which involves commuting the Pauli byproducts from the innermost parentheses to the outermost.

Since the $U_{\vec{C}_i}$ operators are Pauli operators, commuting them through the projectors $P_{\vec{S}_j}$ in Eq.~\eqref{eq: adap} can, at most, flip the signs of the components of ${\vec{S}_j}$
\begin{equation}
\tilde{\rho}_{0,\vec{C}} = \sum_{\vec{S}} (\prod_{\vec{C}_i} U_{\vec{C}_i})P_{\vec{S}'_{t_f}}(\dots(P_{\vec{S}'_2}(P_{\vec{S}'_1}\rho_i P_{\vec{S}'_1})P_{\vec{S}'_2})\dots)P_{\vec{S}'_{t_f}} (\prod_{\vec{C}_i} U_{\vec{C}_i})^{\dagger}\otimes |\vec{S}\rangle\langle\vec{S}|
\end{equation}
By changing the summing variable from $\vec{S}$ to $\vec{S}'$ we get the following
\begin{equation}
\tilde{\rho}_{\vec{C},p} = \sum_{\vec{S}} (\prod_{\vec{C}_i} U_{\vec{C}_i})P_{\vec{S}_{t_f}}(\dots(P_{\vec{S}_2}(P_{\vec{S}_1}\rho_i P_{\vec{S}_1})P_{\vec{S}_2})\dots)P_{\vec{S}_{t_f}} (\prod_{\vec{C}_i} U_{\vec{C}_i})^{\dagger}\otimes |\vec{S}'\rangle\langle\vec{S}'|
\end{equation}
which effectively absorbs the sign changes into the $|\vec{S}\rangle\langle\vec{S}|$ term. The state  $|\vec{S}'\rangle\langle\vec{S}'|$ can be transformed back to state $|\vec{S}\rangle\langle\vec{S}|$ by a unitary transformation which results in the final form of:
\begin{equation}
\tilde{\rho}_{0,\vec{C}} = U_{\vec{C}}\left(\sum_{\vec{S}} P_{\vec{S}_{t_f}}(\dots(P_{\vec{S}_2}(P_{\vec{S}_1}\rho_i P_{\vec{S}_1})P_{\vec{S}_2})\dots)P_{\vec{S}_{t_f}} \otimes |\vec{S}\rangle\langle\vec{S}|\right)U_{\vec{C}}^\dagger.
\end{equation}
where the unitary acts on the code qubits as the product of the $U_{\vec{C}_i}$'s and on the syndromes. 

Consequently, the full resource state is:
\begin{equation}
\rho_0 = \frac{1}{2^{|\vec{C}|}}\sum_{\vec{C}}|\vec{C}\rangle\langle\vec{C}|\otimes U_{\vec{C}}(\tilde{\rho}_0)U_{\vec{C}}^\dagger,
\end{equation}
where $\tilde{\rho}_0$ is the postselected MBQC state equivalent to the circuit state, including syndromes and outputs.

This establishes a quantum channel $F$ relating the MBQC state $\rho_0$ to its equivalent circuit representation $\tilde{\rho}_0$:
\begin{equation}
F(\rho_0) = \tilde{\rho}_0,\quad F^{-1}(\tilde{\rho}_0)=\rho_0,
\end{equation}
explicitly given by:
\begin{equation} \label{eq: F defs}
F(\cdot) = \mathrm{Tr}_C\left(\sum_{\vec{C}}U_{\vec{C}}^\dagger P_{\vec{C}}(\cdot)P_{\vec{C}}U_{\vec{C}}\right),\quad F^{-1}(\cdot)=\frac{1}{2^{|\vec{C}|}}\sum_{\vec{C}}|\vec{C}\rangle\langle\vec{C}|\otimes\left(U_{\vec{C}}(\cdot)U_{\vec{C}}^\dagger\right),
\end{equation}
where $P_{\vec{C}}$ projects onto code qubit measurement outcomes and $U_{\vec{C}}$ are the associated non-adaptive Pauli corrections. Note that $F^{-1}$ serves as a right inverse, satisfying $F \circ F^{-1} = \mathbb{I}$.



\subsection{Feedback Channel for Noisy MBQC} 

We now generalize the relationship in Eq.~\eqref{eq: F defs} to noisy settings. That is, if $\tilde{\rho}_p$ denotes the circuit state with noise parameter $p$ and $\rho_p$ the corresponding noisy MBQC state (defined via the noise map in the main text [Fig.~\ref{fig:Error}]), the feedback channel still holds:
\begin{equation}
F(\rho_p)=\tilde{\rho}_p,\quad F^{-1}(\tilde{\rho}_p)=\rho_p.
\label{Eq:noisy_feed}
\end{equation}

For a noisy circuit, the density matrix is constructed by sequentially applying the measurement channel and the noise channel, expressed as:
\begin{equation}
\tilde{\rho}_p = \left(\circ_{t=1}^{t_f} (M \circ \mathcal{E}_p)\right)(\rho_i),
\end{equation}
where \(\mathcal{E}_p\) denotes the noise channel acting on both the code state and syndrome outcomes, incorporating errors such as \(X\)- and \(Z\)-type noise, as well as readout errors. Explicitly, the noisy state \(\tilde{\rho}_p\) is

\begin{gather}\label{expl_circuit_mixedd}
\tilde{\rho}_p = \sum_{\vec{S}} P_{\vec{S}_{t_f}} \left(\cdots\mathcal{E}_p\left(P_{\vec{S}_2} \left(\mathcal{E}_p\left(P_{\vec{S}_1} \left(\mathcal{E}_p(\rho_i)\right) P_{\vec{S}_1}\right)\right) P_{\vec{S}_2}\right)\cdots\right) P_{\vec{S}_{t_f}} \otimes \mathcal{E}_p(|\vec{S}\rangle\langle\vec{S}|).
\end{gather}

To prove Eq.~\eqref{Eq:noisy_feed}, it suffices to show:
\begin{equation}
F^{-1}(\tilde{\rho}_p)=\frac{1}{2^{|\vec{C}|}} \sum_{\vec{C}} |\vec{C}\rangle\langle\vec{C}| \otimes (U_{\vec{C}} \tilde{\rho}_p U_{\vec{C}}^\dagger) = \rho_p,
\label{eq:main}
\end{equation}
The identity \(F \circ F^{-1} = \mathbb{I}\) then implies the converse: \(
F(\rho_p) = \tilde{\rho}_p
\). 

First, consider the noiseless case (\(p=0\)):
\begin{gather}
\rho_0 = \frac{1}{2^{|\vec{C}|}} \sum_{\vec{C}} |\vec{C}\rangle\langle\vec{C}| \otimes \tilde{\rho}_{0,\vec{C}},
\end{gather}
Applying the \(\mathcal{E}_p^Z\) channel yields:
\begin{gather}
\rho_p = \frac{1}{2^{|\vec{C}|}} \sum_{\vec{C}} \mathcal{E}_p^Z(|\vec{C}\rangle\langle\vec{C}| \otimes \tilde{\rho}_{0,\vec{C}}),
\end{gather}
Let us start by applying the noise channel on only one code qubit (say \(C_1\)), we then have:
\begin{align}
&\mathcal{E}^{C_1}_p\left(\rho_0\right)=\frac{1}{2^{|\vec{C}|}} \sum_{\vec{C}} \left(|\vec{C}'_1\rangle\langle\vec{C}'_1|\right) \otimes \mathcal{E}^z_p\left(|C_1\rangle\langle C_1|\right)\otimes \tilde{\rho}_{0,\vec{C}}\\ &= \frac{1}{2^{|\vec{C}|}} \sum_{\vec{C}} \nonumber\left(|\vec{C}'_1\rangle\langle\vec{C}'_1|\right) \otimes \left((1-p)|C_1\rangle\langle C_1| + p |-C_1\rangle\langle-C_1| \right)\otimes \tilde{\rho}_{0,\vec{C}}\\
&= \frac{1}{2^{|\vec{C}|}} \sum_{\vec{C}} \nonumber|\vec{C}\rangle\langle\vec{C}| \otimes \left((1-p)\tilde{\rho}_{0,\{\vec{C}'_1,C_1\}} + p\; \tilde{\rho}_{0,\{\vec{C}'_1,-C_1\}}\right),
\end{align}
Here, the vector ${\vec{C}}$ is broken into two parts: $\vec{C}'_1$ and $C_1$, where ${\vec{C}'_1}$ denotes all code qubits except \(C_1\). By relabeling the code basis states, the net effect is to move the noise action from the code qubits \(|\vec{C}\rangle\langle\vec{C}|\) onto \(\tilde{\rho}_{0,\vec{C}}\). 

Now, we write the noise on $\tilde{\rho}_{0,\vec{C}}$ explicitly:
\begin{align}
&\left((1-p)\tilde{\rho}_{0,\{\vec{C}'_1,C_1\}} + p\; \tilde{\rho}_{0,\{\vec{C}'_1,-C_1\}}\right) \\
&= \sum_{\vec{S}} P_{\vec{S}_{t_f}} \nonumber\left(...\;U_{\vec{C}_3}\left(P_{\vec{S}_2} \left(U_{\vec{C}_2}\left(P_{\vec{S}_1} U_{\vec{C}_1} \mathcal{E}_p(\rho_i) U_{\vec{C}_1}^\dagger P_{\vec{S}_1}\right)U_{\vec{C}_2}^\dagger\right) P_{\vec{S}_2}\right)U_{\vec{C}_3}^\dagger...\right) P_{\vec{S}_{t_f}} \otimes (|\vec{S}\rangle\langle\vec{S}|),
\label{eq:alomsttt}
\end{align}
where the noise channel $\mathcal{E}_p$ is $\mathcal{E}_p(\cdot) = (1-p)(\cdot ) + p\; U_{C_1}U_{-C_1}(\cdot)U_{-C_1}U_{C_1}$, with the byproduct operators $U_{C_1},U_{-C_1}$ in Eq.~\eqref{eq: byproduct operators}. Note that the factor of $U_{C_1}$ in $U_{\vec{C}_1}$ cancels with the factor of $U_{C_1}$ in the action of $\mathcal{E}_p$.
Extending this procedure to all qubits yields the explicit form:
\begin{gather}
\rho_p = \frac{1}{2^{|\vec{C}|}} \sum_{\vec{C}} |\vec{C}\rangle\langle\vec{C}| \otimes \tilde{\rho}_{\vec{C},p},\\
\tilde{\rho}_{\vec{C},p} = \sum_{\vec{S}} P_{\vec{S}_{t_f}} \left(...\;U_{\vec{C}_3}\mathcal{E}_p\left(P_{\vec{S}_2} \left(U_{\vec{C}_2}\mathcal{E}_p\left(P_{\vec{S}_1} (U_{\vec{C}_1} \mathcal{E}_p(\rho_i) U_{\vec{C}_1}^\dagger)  P_{\vec{S}_1}\right)U_{\vec{C}_2}^\dagger\right) P_{\vec{S}_2}\right)U_{\vec{C}_3}^\dagger...\right) P_{\vec{S}_{t_f}} \otimes \mathcal{E}_p(|\vec{S}\rangle\langle\vec{S}|) 
\end{gather}

Now we aim to commute the $U_{\vec{C}_i}$ operators through the \(\mathcal{E}_p\)'s and the \(P_{\vec{S}_i}\)'s, moving it from inside the parentheses to the outermost position. First, note that \(U_{\vec{C}_i}\) commutes with any \(\mathcal{E}_p\), or equivalently: 
\(
\mathcal{E}_p(U_{\vec{C}_i}(\cdot) U_{\vec{C}_i}^\dagger) = U_{\vec{C}_i}\mathcal{E}_p(\cdot)U_{\vec{C}_i}^\dagger.
\)
Second, \(U_{\vec{C}_j}\) either commutes with \(P_{\vec{S}_i}\) or changes the sign of \(S_i\). We define $\vec{S}'_i$ according to:
\(
P_{\vec{S}_i}(U_{\vec{C}_j}(\cdot)U_{\vec{C}_j}^\dagger)P_{\vec{S}_i}^\dagger = U_{\vec{C}_j}(P_{\vec{S}'_i}(\cdot)P_{\vec{S}'_i}^\dagger)U_{\vec{C}_j}^\dagger.
\)
Using these two commutation relations, we obtain:
\begin{equation}
\tilde{\rho}_{\vec{C},p} = \sum_{\vec{S}} (\prod_{C_i} U_{C_i})P_{\vec{S}'_{t_f}} \left(...\;\mathcal{E}_p\left(P_{\vec{S}'_2} \left(\mathcal{E}_p\left(P_{\vec{S}'_1} \mathcal{E}_p(\rho_i) P_{\vec{S}'_1}\right)\right) P_{\vec{S}'_2}\right)...\right) P_{\vec{S}'_{t_f}}(\prod_{C_i} U_{C_i})^{\dagger} \otimes \mathcal{E}_p(|\vec{S}\rangle\langle\vec{S}|) 
\end{equation}
Given that \(|\vec{S}\rangle\langle\vec{S}|\) can be expressed as a correction applied to \(|\vec{S}'\rangle\langle\vec{S}'|\) with a unitary feedback that flips the relevant syndromes, we obtain:
\begin{gather}\label{almostend}
\tilde{\rho}_{\vec{C},p} = U_{\vec{C}}\left(\sum_{\vec{S}'} P_{\vec{S}'_{t_f}} \left(...\;\mathcal{E}_p\left(P_{\vec{S}'_2} \left(\mathcal{E}_p\left(P_{\vec{S}'_1} \mathcal{E}_p(\rho_i) P_{\vec{S}'_1}\right)\right) P_{\vec{S}'_2}\right)...\right) P_{\vec{S}'_{t_f}} \otimes \mathcal{E}_p(|\vec{S}'\rangle\langle\vec{S}'|) 
\right)U_{\vec{C}}^{\dagger}.
\end{gather}

Inserting Eq.~\eqref{almostend} into Eq.~\eqref{expl_circuit_mixedd}, we obtain:
\begin{align}
    \tilde{\rho}_{\vec{C},p} = U_{\vec{C}} \tilde{\rho}_p U_{\vec{C}}^{\dagger}.
\end{align}
Finally, combining this with Eq.~\eqref{eq:alomsttt}, we arrive at:
\begin{equation}
\rho_p = \frac{1}{2^{|\vec{C}|}} \sum_{\vec{C}} |\vec{C}\rangle\langle\vec{C}| \otimes (U_{\vec{C}} \tilde{\rho}_p U_{\vec{C}}^{\dagger}),
\end{equation}
which completes the proof.




\section{Coherent errors in the syndrome extraction circuit}
\label{app:B}

In the main text, we established a connection between the threshold of the syndrome extraction circuit and a mixed-state phase transition under incoherent noise.
In this appendix, we consider instead coherent errors in the syndrome extraction circuit and use two complementary approaches. We first consider the \(F\) channel from Appendix~\ref{app:A} and discuss its subtleties and limitations in the presence of coherent errors. We nevertheless demonstrate that for the simple case of the repetition code with coherent errors acting only on the code qubits, this formalism successfully establishes a connection between mixed-state phase transitions and the fault-tolerance threshold.

Next, we use measurement and postselection on the resource state to show that the CMI of the syndromes in the circuit is equal to that in the MBQC model in the presence of coherent errors. This result is particularly useful, as it enables a correspondence whereby the divergence of the Markov length in one setup implies the divergence of the Markov length in the other. However, it is not yet strong enough to establish that the threshold under coherent noise corresponds to a mixed-state phase transition.

\subsection{Coherent error and F channel}
Here we study whether the fault-tolerance transition observed in a syndrome extraction circuit subjected to coherent errors can be mapped to a mixed-state phase transition of the $X$-dephased resource state. To establish this correspondence using the arguments in the main text, we must demonstrate that the mixed state of the circuit under coherent errors, denoted by $\tilde{\rho}_\theta$, maps via the inverse transformation $F^{-1}(\tilde{\rho}_\theta)$ to a local Lindbladian evolution acting upon a completely decohered resource state $\rho_0$: 
\begin{equation}
    F^{-1}(\tilde{\rho}_\theta) = \mathcal{T} e^{\int_0^{t_p} \mathcal{L}_1(t) dt}[\rho_0].
\label{Eq:coh_lind}
\end{equation}
For simplicity, we consider the case where the coherent noise acts only on the code qubits.

We first demonstrate explicitly how $\tilde{\rho}_\theta$ decomposes into a combination of correlated incoherent errors and a residual term that we expect to vanish in the thermodynamic limit. We show vanishing of the residual term explicitly for the repetition code syndrome extraction circuit. Subsequently, using the transformation $F^{-1}$, correlated incoherent errors in the circuits maps to correlated incoherent errors within the resource state. We further verify explicitly in the repetition-code case that such errors correspond precisely to a local Lindbladian evolution. However, the generality of this construction beyond the repetition-code setting remains an open question.\\

The evolution of the density matrix is modeled as a sequence of alternating noise and measurement channels:
\begin{equation}
    \rho_f = \left( \circ_{t=1}^{t_f} (M \circ \mathcal{E}) \right)(\rho_i),
    \label{eq:circuit_channel}
\end{equation}
where \(\mathcal{E}\) denotes the noise channel that acts with coherent noise on the code qubits and readout errors on the syndrome outcomes. The final state of the system, after extracting syndromes \(\vec{S} = (\vec{S}_1, \vec{S}_2, \dots, \vec{S}_{t_f})\), is given by:
\begin{equation}
    \rho_f = \sum_{\vec{S}} P_{\vec{S}_{t_f}} \left(\cdots \mathcal{E}\left(P_{\vec{S}_2} \left(\mathcal{E}\left(P_{\vec{S}_1} \left(\mathcal{E}(\rho_i)\right) P_{\vec{S}_1}\right)\right) P_{\vec{S}_2}\right)\cdots \right) P_{\vec{S}_{t_f}} \otimes \mathcal{E}_p(|\vec{S}\rangle\langle\vec{S}|),
    \label{eq:explicit_mixed_state}
\end{equation}
where \(P_{\vec{S}_t}\) are the syndrome projection operators and \(\mathcal{E}_p\) is the readout error channel on the classical syndrome record. 

From here onward, we assume the noise channel \(\mathcal{E}\) acts as a coherent unitary rotation \(e^{i \alpha \hat{X}} = \cos(\alpha) + i \sin(\alpha)\hat{X}\) on code qubits and the underling code is a stabilizer code. The action of this channel on the initial state \(\rho_i\) can be expanded as:
\begin{equation}
    \mathcal{E}(\rho_i) = \sum_{\vec{m}} C_{\vec{m}} \hat{X}^{\vec{m}_L} \rho_i \hat{X}^{\vec{m}_R},
    \label{eq:coherent_noise_expansion}
\end{equation}
where \(\vec{m}_L\) and \(\vec{m}_R\) denote binary vectors indexing Pauli-\(X\) operators on each qubit, therfore $\hat{X}^{\vec{m}} = \hat{X}^{\vec{m}_1}\hat{X}^{\vec{m}_2} \cdots \hat{X}^{\vec{m}_n}$. \(C_{\vec{m}}\) are coefficients determined by products of \(\cos(\alpha)\) and \(i \sin(\alpha)\). Now we move to the next parenthesis of Eq.~\eqref{eq:explicit_mixed_state} which projects to the syndrome outcome \(\vec{S}_1\):
\begin{equation}
    P_{\vec{S}_1} \mathcal{E}(\rho_i) P_{\vec{S}_1} = \sum_{\vec{m}} C_{\vec{m}} P_{\vec{S}_1} \hat{X}^{\vec{m}_L} \rho_i \hat{X}^{\vec{m}_R} P_{\vec{S}_1}.
    \label{eq:projected_noise}
\end{equation}
This can be simplified taking the initial state to be \(\rho_i = \rho_L P_{\vec{S}=0}\), where $P_{\vec{S}=0}$ projects in to the code subspace and $\rho_L$ further chooses a specific code state:
\begin{equation}
    P_{\vec{S}_1} \hat{X}^{\vec{m}_L} \rho_i \hat{X}^{\vec{m}_R} P_{\vec{S}_1} = P_{\vec{S}_1} \hat{X}^{\vec{m}_L} \rho_L P_{\vec{S}=0} \hat{X}^{\vec{m}_R} P_{\vec{S}_1} = \rho'_L P_{\vec{S}_1} P_{\vec{S}_0} \hat{X}^{\vec{m}_L} \hat{X}^{\vec{m}_R} P_{\vec{S}_1},
    \label{eq:symmetry_commute}
\end{equation}
where \(P_{\vec{S}_0} = \hat{X}^{\vec{m}_L}P_{\vec{S}=0}\hat{X}^{\vec{m}_L}\) and \(\rho_L' = \hat{X}^{\vec{m}_L}\rho_L\hat{X}^{\vec{m}_L}\). Now note that if \(\hat{X}^{\vec{m}_L} \hat{X}^{\vec{m}_R}\) does not commute with \(P_{\vec{S}_1}\), the \(P_{\vec{S}_1} P_{\vec{S}_0} \hat{X}^{\vec{m}_L} \hat{X}^{\vec{m}_R} P_{\vec{S}_1}\) will be zero. This is because by defining $P_{\vec{S}'_1} = \hat{X}^{\vec{m}_L} \hat{X}^{\vec{m}_R} P_{\vec{S}_1}\hat{X}^{\vec{m}_L} \hat{X}^{\vec{m}_R}$ the following equation holds since the projections are orthogonal.
\begin{equation}
    P_{\vec{S}_1} P_{\vec{S}_0} P_{\vec{S}'_1} = 0,
    \label{eq:orthogonal_projectors}
\end{equation}

Thus, only terms for which \(\hat{X}^{\vec{m}_L} \hat{X}^{\vec{m}_R}\) commutes with \(P_{\vec{S}_1}\) survive. These terms correspond to logical operators or stabilizers. We can decompose the sum in Eq.~\eqref{eq:projected_noise} into three contributions: 
\begin{align}
    P_{\vec{S}_1} \mathcal{E}(\rho_i) P_{\vec{S}_1}^\dagger = &\sum_{\vec{m}_L} C^1_{\vec{m}_L} P_{\vec{S}_1} \hat{X}^{\vec{m}_L} \rho_i \hat{X}^{\vec{m}_L} P_{\vec{S}_1} \nonumber \\
    &+ \sum_{S}\sum_{\vec{m}_L} C^2_{\vec{m}_L} P_{\vec{S}_1} \hat{X}^{\vec{m}_L} \rho_i X^S \hat{X}^{\vec{m}_L} P_{\vec{S}_1} \nonumber \\
    &+ \sum_{\vec{m}_L} D_{\vec{m}_L} P_{\vec{S}_1} \hat{X}^{\vec{m}_L} \rho_i \hat{X}^{\vec{m}_L} \bar{X} P_{\vec{S}_1},
    \label{eq:three_term_decomposition}
\end{align}
where \(X^S\) are stabilizers and \(\bar{X}\) denotes a logical Pauli-\(X\) operator. Since \(\rho_i\) is stabilized by \(X^S\), the second term can be absorbed into the first. We are left with:
\begin{align}
    P_{\vec{S}_1} \mathcal{E}(\rho_i) P_{\vec{S}_1}^\dagger = &\sum_{\vec{m}_L} C_{\vec{m}_L} P_{\vec{S}_1} \hat{X}^{\vec{m}_L} \rho_i \hat{X}^{\vec{m}_L} P_{\vec{S}_1} \nonumber \\
    &+ \sum_{\vec{m}_L} D_{\vec{m}_L} P_{\vec{S}_1} \hat{X}^{\vec{m}_L} \rho_i \hat{X}^{\vec{m}_L} \bar{X} P_{\vec{S}_1}.
    \label{eq:sym_asym_split}
\end{align}

We define maps as the following:
\begin{align}
    \mathcal{N}_{\text{sym}}(\rho_i) &= \sum_{\vec{m}_L} C_{\vec{m}_L} \hat{X}^{\vec{m}_L} \rho_i \hat{X}^{\vec{m}_L}, \\
    \mathcal{N}_{\text{asym}}(\rho_i) &= \sum_{\vec{m}_L} D_{\vec{m}_L} \hat{X}^{\vec{m}_L} \rho_i \hat{X}^{\vec{m}_L} \bar{X}.
    \label{eq:channel_defs}
\end{align}
Thus, the total channel can be expressed as:
\begin{equation}
    P_{\vec{S}_1} \mathcal{E}(\rho_i) P_{\vec{S}_1} = P_{\vec{S}_1} \left( \mathcal{N}_{\text{sym}}(\rho_i) + \mathcal{N}_{\text{asym}}(\rho_i) \right) P_{\vec{S}_1}.
    \label{eq:final_channel_decomp}
\end{equation}

The symmetric part, \(\mathcal{N}_{\text{sym}}\), corresponds to correlated incoherent noise and can be mapped to a corresponding error model in MBQC using the construction introduced in the main text. The asymmetric part, \(\mathcal{N}_{\text{asym}}\), does not follow this mapping. However, if one is interested only in mixed-state properties—which should be independent of the logical encoding—one can either select a suitable \(\rho_i\) to further symmetrize the asymmetric term or show that \(\mathcal{N}_{\text{asym}}\) is suppressed in the thermodynamic limit. In the next subsection, we explicitly demonstrate this suppression for the case of the repetition code. And we further show the Eq.~\eqref{Eq:coh_lind} holds for the repetition code by showing that $\mathcal{N}_{\text{asym}}$ is a local channel, however the locality of this channel for the general setting remains an open question.

\subsection{Coherent error in repetition code} 

In the case of the repetition code, $\hat{X}^{\vec{m}_L}$ and $\hat{X}^{\vec{m}_R}$ must satisfy \(\hat{X}^{\vec{m}_L} \hat{X}^{\vec{m}_R} = I \text{\;or\;} \bar{X}\), since there is no \(X\)-type stabilizer. This simplifies the following sum:
\begin{equation}
P_{\vec{S}_1} \left(\mathcal{E}(\rho_i)\right) P_{\vec{S}_1}^\dagger = \sum_{\vec{m}_L} C^1_{\vec{m}_L} P_{\vec{S}_1} \left(\hat{X}^{\vec{m}_L} (\rho_i) \hat{X}^{\vec{m}_L}\right) P_{\vec{S}_1} + \sum_{\vec{m}_L} C^2_{\vec{m}_L} P_{\vec{S}_1} \left(\hat{X}^{\vec{m}_L} (\rho_i) \hat{X}^{\vec{m}_L} \bar{X}\right) P_{\vec{S}_1}.
\label{Eq:symasym}
\end{equation}
Here, \(C^1_{\vec{m}_L} = (\sin^2\alpha)^{|m_L|}(\cos^2\alpha)^{n-|m_L|}\) and \(C^2_{\vec{m}_L} = (i\sin{\alpha} \cos{\alpha})^n (-1)^{|m_L|}\). 


The first term represents symmetric contributions and can be interpreted as incoherent errors. We write the first and the second terms as: 
\begin{equation}
P_{\vec{S}_1} \left(\mathcal{E}(\rho_i)\right) P_{\vec{S}_1} = P_{\vec{S}_1} \left(\mathcal{N}_{\text{sym}}(\rho_i) + \mathcal{N}_{\text{asym}}(\rho_i)\right) P_{\vec{S}_1}
\end{equation}
This implies that the full channel can be written as:
\begin{gather}\label{expl_circuit_mixed}
\left( \circ_{t=1}^{t_f} (M \circ \mathcal{E}) \right)(\rho_i) = \left( \circ_{t=1}^{t_f} (M \circ \left(\mathcal{N}_{\text{sym}} + \mathcal{N}_{\text{asym}}\right)) \right)(\rho_i) 
\end{gather}

Now we want to show that the full noise channel can be approximated by the local dephasing channel as:
\begin{equation}
    \left
( \circ_{t=1}^{t_f} (M \circ \mathcal{E}) \right)(\rho_i) \approx  \left( \circ_{t=1}^{t_f} (M \circ \mathcal{E}_p) \right)(\rho_i)
\end{equation}

Let us first define the following density matrix: 
\begin{gather}\label{expl_circuit_mixed}
\rho_t = \left( \circ_{t+1}^{t_f} (M \circ \mathcal{E}) \right)\circ\left( \circ_{t=1}^{t} (M \circ \mathcal{N}_{\text{sym}} ) \right)(\rho_i) 
\end{gather}
This is the density matrix of the state evolved to time \( t \), undergoing only the \( \mathcal{N}_\text{sym} \) noise, and subsequently experiencing the \( \mathcal{E} \) noise at each time step beyond \( t \). Using a telescoping sum, we have:
\begin{equation}
\sum_{t=0}^{t=t_f} \rho_{t+1} -\rho_t =  \left( \circ_{t=1}^{t_f} (M \circ \mathcal{N}_{\text{sym}})\right)(\rho_i) -  \left( \circ_{t=1}^{t_f} (M \circ \mathcal{E}) \right)(\rho_i)
\end{equation}
This implies:
\begin{equation}
||\left( \circ_{t=1}^{t_f} (M \circ \mathcal{N}_{\text{sym}})\right)(\rho_i) - \left( \circ_{t=1}^{t_f} (M \circ \mathcal{E}) \right)(\rho_i)||_1 \leq \sum_{t=0}^{t=t_f} ||\rho_{t+1} -\rho_t||_1
\end{equation}

Now let's bound each \(||\rho_{t+1} -\rho_t||_1\):
\begin{equation}
    \rho_{t+1} -\rho_t = \left( \circ_{t+2}^{t_f}( M \circ \mathcal{E}) \right)\circ\left(  M \circ \mathcal{N}_{\text{sym}} - M \circ \mathcal{E} \right)\circ(\circ_{t=1}^{t} (M \circ \mathcal{N}_{\text{sym}}))(\rho_i) 
\end{equation}
Using the data processing inequality we have:
\begin{equation}
    ||\rho_{t+1} -\rho_t||_1 \leq ||(M \circ \mathcal{N}_{\text{sym}}- M \circ \mathcal{E})(\bar{\rho})||_1 = ||(M \circ \mathcal{N}_{\text{asym}})(\bar{\rho})||_1
\end{equation}
where \(\bar{\rho} = (\circ_{t=1}^{t} (M \circ \mathcal{N}_{\text{sym}}))(\rho_i) \). Form the Eq.~\eqref{Eq:symasym} we have:
\begin{equation}
    (M \circ \mathcal{N}_{\text{sym}}- M \circ \mathcal{E})(\bar{\rho}) = \sum_{\vec{S}_1}\sum_{\vec{m}_L} C^2_{\vec{m}_L} P_{\vec{S}_1} \left(\hat{X}^{\vec{m}_L} (\bar{\rho}) \hat{X}^{\vec{m}_L} \bar{X}\right) P_{\vec{S}_1}\otimes |\vec{S}_1\rangle\langle\vec{S}_1|
\end{equation}
Bounding the right-hand side and again using data processing inequality:
\begin{equation}
    ||\sum_{\vec{S}_1}\sum_{\vec{m}_L} C^2_{\vec{m}_L} P_{\vec{S}_1} \left(\hat{X}^{\vec{m}_L} (\bar{\rho}) \hat{X}^{\vec{m}_L} \bar{X}\right) P_{\vec{S}_1}\otimes |\vec{S}_1\rangle\langle\vec{S}_1||| \leq \sum_{\vec{m}_L} |C^2_{\vec{m}_L} | \;|| \left(\hat{X}^{\vec{m}_L} (\bar{\rho}) \hat{X}^{\vec{m}_L} \bar{X}\right)||
\end{equation}
Noting that \(|| \left(\hat{X}^{\vec{m}_L} (\bar{\rho}) \hat{X}^{\vec{m}_L} \bar{X}\right)||_1 = ||\bar{\rho}||_1 = 1\), we get:
\begin{equation}
||\rho_{t+1} -\rho_t||_1 \leq \sum_{\vec{m}_L} |C^2_{\vec{m}_L} | = (2\sin{\alpha} \cos{\alpha})^n
\end{equation}
This finally results in:
\begin{gather}
||\left( \circ_{t=1}^{t_f} (M \circ \mathcal{E}) \right)(\rho_i) -  \left( \circ_{t=1}^{t_f} (M \circ \mathcal{N}_{\text{sym}})\right)(\rho_i) ||_1 \leq n \; (\sin(2\alpha))^n.
\end{gather}
This term will vanish as long as \(\sin(2\alpha)\leq 1\). Now, let us define the Pauli-twirled probability \(p = \sin^2{\alpha}\). Then, we have \(C^1_{\vec{m}_L} = (p)^{|m_L|}(1-p)^{n-|m_L|}\), which results in:
\begin{equation}
    \left( \circ_{t=1}^{t_f} (M \circ \mathcal{N}_{\text{sym}})\right)(\rho_i) = \left( \circ_{t=1}^{t_f} (M \circ \mathcal{E}_p) \right)(\rho_i) 
\end{equation}
where \(\mathcal{E}_p\) is a dephasing channel with probability \(p\). This finally establishes that:
\begin{gather}
||\left( \circ_{t=1}^{t_f} (M \circ \mathcal{E}) \right)(\rho_i) -  \left( \circ_{t=1}^{t_f} (M \circ \mathcal{E}_p) \right)(\rho_i) ||_1 \leq n \; (\sin(2\alpha))^n
\end{gather}
Which is exponentially small as long as \(\sin^2{\alpha} = p\leq \frac{1}{2}\), thereby completing the proof.

\subsection{Coherent error and forced measurements}
\label{app:mbqc-syndrome}

Here, we establish an explicit mapping (via postselection) between the density matrix of the \(X\)-decohered noisy resource state and the corresponding syndrome distribution in the circuit model. This allows us to relate the CMI of the syndromes in a circuit subjected to coherent error to that of a resource state subjected to the corresponding coherent error.

The mapping that we introduce can be interpreted as a form of post-selection. We define a post-selection map \( \mathcal{P} \) by
\begin{equation}
    \mathcal{P}(\rho_\theta) = \frac{\Pi^C \rho_{\theta} \Pi^C}{\operatorname{Tr}[\Pi^C \rho_\theta]},
    \label{eq:postselection}
\end{equation}
where \( \Pi^C \) projects all code qubits onto the \( \ket{+} \) state. In words, we measure all of the code qubits of the resource state and only keep the result with $|+\rangle$ on every site.

In the context of the main text, we show that this post-selection yields the syndrome density matrix:
\begin{equation}
    \mathcal{P}(\rho_\theta) = \bar{\rho}_\theta,
    \label{eq:syndrome-map}
\end{equation}
where \( \rho_\theta \) is the density matrix of the \(X\)-decohered noisy resource state, and \( \bar{\rho}_\theta \) represents the syndrome distribution in the corresponding circuit.

\begin{figure}[h!]
    \centering
    \includegraphics[width=0.6\linewidth]{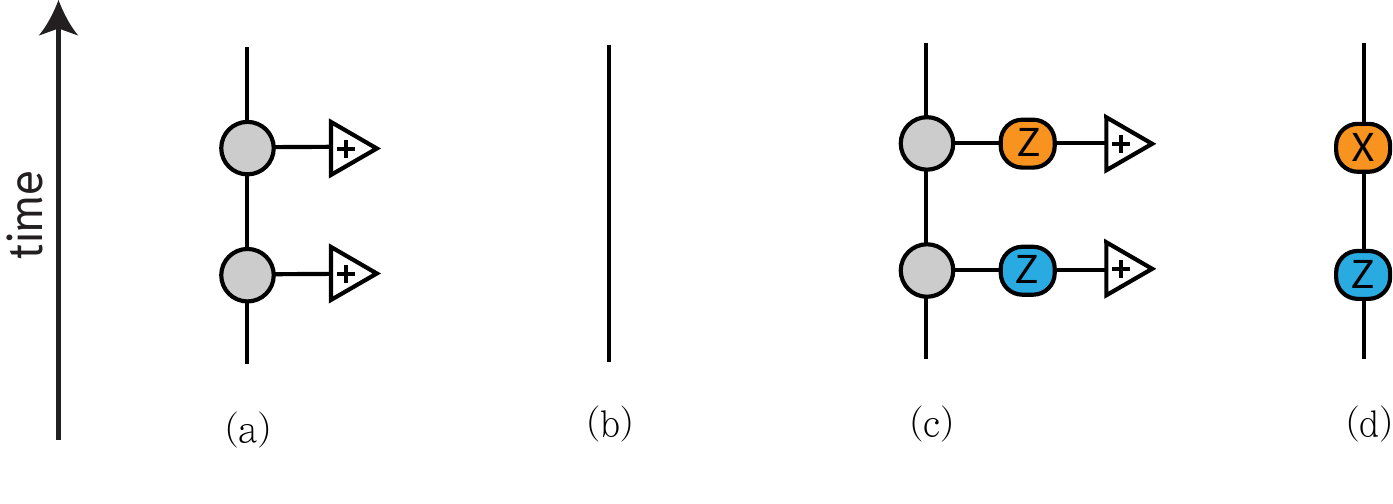}
    \caption{
    (a) Code qubits at times \( t \) and \( t + \frac{1}{2} \) are post-selected into the \( \ket{+} \) state, indicated by contraction with a triangle symbol. 
    (b) This post-selection implements perfect teleportation, effectively enacting a time evolution from time \( t \) to \( t + 1 \). 
    (c) A coherent error unitary is applied to the code qubits, represented by the blue and orange gates. The post-selection is performed afterward. The Kraus operators of both noise channels are constructed using only Pauli \( Z \). 
    (d) The net effect is equivalent to applying the same coherent error during teleportation. In the circuit picture, the blue channel is applied directly to the teleported qubit, while the orange channel is mapped to an equivalent channel in which each Pauli \( Z \) operator is replaced by a Pauli \( X \).
    }
    \label{fig:error-mapping}
\end{figure}

To validate Eq.~\eqref{eq:syndrome-map} and establish the correspondence, we analyze the effect of forced measurements on a code qubit subject to noise. We demonstrate that such a measurement effectively transfers the error to the teleported qubit, thereby implementing the same noise process. Specifically, we consider the teleportation of qubit \( i \) from time \( t \) to time \( t+1 \), realized via measurements at time slices \( t \) and \( t + \frac{1}{2} \), as described in Refs.~\cite{Brown_2020,hauser2024informationdynamicsdecoheredquantum}. The net effect is to propagate the quantum state to time \( t+1 \), with the error carried along the trajectory, as shown in Fig.~\ref{fig:error-mapping}.

We do not provide a direct proof of the correspondence illustrated in Fig.~\ref{fig:error-mapping}, as the argument is straightforward and involves applying gates and measurements on a three-qubit system. Given this correspondence, we have that post-selecting the code qubits precisely implements the circuit model, since the syndrome qubits are extracted identically in both the MBQC and circuit frameworks.

Having established this mapping, we now demonstrate that it preserves the CMI. In particular, this shows that a divergence of Markov length in one framework implies a corresponding divergence in the other. The proof proceeds in two steps. First, we show that the CMI associated with the detector cells is preserved under the mapping and coincides with that of the deterministic components in the circuit. Second, we use the argument from the main text, namely that the total CMI depends solely on the contributions from the detector cells, which guarantees the preservation of CMI across the two setups.

The post-measurement distribution of the noisy resource state can be expressed as a joint distribution over detector cells and random components. In the noiseless case, measurement outcomes are entirely random except for those associated with detector cells. This is reflected in the expectation values:
\[
\left \langle \prod_{i \in D} X_i \right \rangle = 1,\quad \left \langle \prod_{j \in D'} X_j \right \rangle = 0 \;\;\; \text{ for } D' \text{ not a detector cell}.
\]

This can be shown for noiseless resource state by identifying a stabilizer of the resource state of the form \( P_{j'} \prod Z_j \), where \( P_{j'} \) is a Pauli operator supported outside \( D' \), and \( \prod Z_j \) is the only component overlapping with \( D' \). The resulting anticommutation between \( \prod Z_j \) and \( \prod_{j \in D'} X_j \) ensures that the expectation value \( \langle \prod_{j \in D'} X_j \rangle \) vanishes in the resource state.

In the presence of coherent noise, the condition \( \langle \prod_{j \in D'} X_j \rangle = 0 \) remains valid for the cluster state, as well as any other resource state admitting a stabilizer of the form \( P_{j'} \prod Z_j \), with support disjoint from \( D' \) except for the \( Z \)-type terms. This ensures that the random components retain their full randomness.

To make this precise, consider the ideal resource state \( \rho_0 \), stabilized by such an operator. When an onsite coherent \( Z \)-type error is applied, the resulting state \( \rho_\theta \) is stabilized by the conjugated operator \( O_{j'} \prod Z_j \), where \( O_{j'} \) acts exclusively on qubits outside of \( D' \). Importantly, the \( \prod Z_j \) term remains unaffected by the error, and hence continues to anticommute with \( \prod_{j \in D'} X_j \). As a result, the expectation value 
\(
\left\langle \prod_{j \in D'} X_j \right\rangle = 0
\)
still holds in the noisy state, confirming that measurement outcomes on non-detector cells remain fully random.


This decomposition into detector cells and random components allows us to express the state as
\begin{equation}
    \rho_\theta = \mathcal{E}_{p=\frac{1}{2}}^x \big( \mathcal{E}^z_\theta(\rho_{\text{RS}}) \big) = \sum_{m, c, s} P_m\; \delta(c,s \equiv m) \ket{c, s}\bra{c, s},
\end{equation}
where \( P_m \) denotes the probability of observing the detector outcome \( m \). The state \( \ket{c, s} \) labels the measurement outcomes of the code qubits (\( c \)) and syndrome qubits (\( s \)). The delta function \( \delta(c, s \equiv m) \) enforces consistency between the code and syndrome configurations with the expected detector outcome \( m \).

Applying post-selection via \( \Pi^C \), we obtain 
\begin{equation}
    \bar{\rho}_\theta \propto \Pi^C \rho_\theta \, \Pi^C = \sum_{m,c, s} P_m\; \delta(c,s \equiv m) \Pi^C \ket{c, s}\bra{c, s} \Pi^C \propto \sum_{s,m} P_m\; \delta(s \equiv m) \ket{s}\bra{s},
\end{equation}
where \( \ket{s} \propto \Pi^C \ket{c, s} \) are the normalized projections onto a fixed configuration of the code qubits, which we suppress for simplicity. Since each detector cell acts on both code and syndrome qubits, fixing the code configuration ensures consistency with the corresponding syndrome outcomes, which is enforced by the constraint \( \delta(s \equiv m) \).

On the other hand, the syndrome distribution in the circuit model can be decomposed as
\begin{equation}
    \bar{\rho}_\theta = \sum_{s,d} P_d\; \delta(s \equiv d) \ket{s}\bra{s},
\end{equation}
where \( d \) labels the deterministic components of the circuit.

By comparing the two expressions, we conclude that the probabilities are proportional. Given the one-to-one correspondence between detector cells in MBQC and deterministic components in the circuit, this proportionality implies equality:
\begin{equation}
    P_{d_m} = P_m,
\end{equation}
establishing that the probabilities associated with detector outcomes in MBQC match those of the corresponding deterministic components in the circuit model.

As discussed in the main text, the entropy of any region in both the circuit and the resource state can be decomposed into contributions from the deterministic (or detector) part and the random part. Since the random part does not contribute to the CMI, this equivalence of probabilities implies that the CMI is identical across the MBQC and circuit models, independent of the nature of the underlying noise.

\section{Coherent error in MBQC}
\label{app:C}

In the previous appendix, we were unable to find a simple form for the noise on the mixed state given coherent noise in the circuit model, except in the case of the repetition code. Here, we ask whether there is an associated mixed-state phase transition, given coherent noise acting on the MBQC resource state. We find that, indeed, coherent noise on the resource state maps to iid incoherent noise on the mixed state. This establishes a connection between the fault-tolerance threshold of MBQC to coherent noise and a mixed-state phase transition. In particular, we show that there exists a channel \( \mathcal{G} \) that maps MBQC with coherent errors to MBQC with incoherent errors:
\begin{eqs}
    \mathcal{G}\left(\mathcal{E}^x_{p=1/2} \left( \mathcal{E}^z_{\theta}(\rho_{\mathrm{RS}}) \right)\right) = \mathcal{E}^x_{p=1/2} \left( \mathcal{E}^z_{p}(\rho_{\mathrm{RS}}) \right)
\end{eqs}
Where $\mathcal{E}^z_{p}$ is the Pauli-twirled of $\mathcal{E}^z_{\theta}$. This shows that the existence of a recovery channel for incoherent errors implies the existence of a recovery channel for coherent errors in MBQC.


To show the existence of $\mathcal{G}$, we begin by defining $\rho_\theta$ as the state obtained by applying coherent \(Z\)-errors, followed by maximal \(X\)-dephasing to the resource state:
\begin{equation}
\rho_{\theta} = \mathcal{E}^x_{p=1/2} \left( \mathcal{E}^z_{\theta}(\rho_{\mathrm{RS}}) \right)
\label{eq:rho_theta}
\end{equation}
Here, \(\mathcal{E}^z_{\theta}\) denotes a coherent rotation error, and \(\mathcal{E}^x_{p=1/2}\) denotes a maximally dephasing \(X\)-error channel. To analyze the behavior under the feedback channel \(F\), we introduce the measurement channel \(\mathcal{E}_D^x\) representing the measurement of the detector cells. Note that

\begin{equation}
\mathcal{E}^x_{p=1/2} = \mathcal{E}^x_{p=1/2} \circ \mathcal{E}_D^x.
\end{equation}

We define the intermediate state
\begin{equation}
\rho'_{\theta} = \mathcal{E}_D^x \left( \mathcal{E}^z_{\theta}(\rho_{\mathrm{RS}}) \right),
\label{eq:rho_prime}
\end{equation}
so that \(\rho_{\theta} = \mathcal{E}^x_{p=1/2}(\rho'_{\theta})\).

Expanding the coherent error, we have:
\begin{equation}
\mathcal{E}_D^x \left( \mathcal{E}^z_{\theta}(\rho_{\mathrm{RS}}) \right) 
= \sum_{\vec{D}} \sum_{\vec{m}} C_{\vec{m}} P_{\vec{D}} \hat{Z}^{\vec{m}_L} \rho_{\mathrm{RS}} \hat{Z}^{\vec{m}_R} P_{\vec{D}}.
\label{eq:ED_expansion}
\end{equation}
The terms in this expansion vanish unless
\begin{equation}
\hat{Z}^{\vec{m}_L} \hat{Z}^{\vec{m}_R} P_{\vec{D}}  =  P_{\vec{D}} \hat{Z}^{\vec{m}_L} \hat{Z}^{\vec{m}_R},
\end{equation}
So we take \( \hat{Z}^{\vec{m}_R} = \hat{Z}^{\vec{m}_{LR}} \hat{Z}^{\vec{m}_L}\). 
\begin{equation}
\mathcal{E}^x_{p=1/2} \left( \mathcal{E}^z_{\theta}(\rho_{\mathrm{RS}}) \right) = 
\mathcal{E}^x_{p=1/2} \left[ \sum_{\vec{m}_L,\vec{m}_{LR}} C_{\vec{m}} \hat{Z}^{\vec{m}_L} \rho_{\mathrm{RS}} \hat{Z}^{\vec{m}_{LR}}\hat{Z}^{\vec{m}_L} \right],
\label{eq:coherent_sum}
\end{equation}

We now want to study  
\(
F\left( \mathcal{E}^x_{p=1/2}(\mathcal{E}^z_{\theta}(\rho_{\mathrm{RS}})) \right).
\) 
First, note that since \(\hat{Z}^{\vec{m}_{LR}}\) commutes with the detector cells—which are all constructed as products of Pauli \(X\) operators—there must exist an operator \(\hat{X}^{\vec{n}_{LR}}\) such that \(\hat{Z}^{\vec{m}_{LR}}\hat{X}^{\vec{n}_{LR}}\) is a stabilizer of the resource state. In other words, the resource state is stabilized by \(\hat{Z}^{\vec{m}_{LR}}\hat{X}^{\vec{n}_{LR}}\). We can therefore write:
\begin{eqs}
    F\left( \mathcal{E}^x_{p=1/2}(\mathcal{E}^z_{\theta}(\rho_{\mathrm{RS}})) \right) =\sum_{\vec{m}_L,\vec{n}_{LR}} C'_{\vec{m}} F\left( \mathcal{E}^x_{p=1/2}(\hat{Z}^{\vec{m}_L} \rho_{\mathrm{RS}} \hat{Z}^{\vec{m}_L} \hat{X}^{\vec{n}_{LR}})) \right)
    \label{Eq:F_channel_asym}
\end{eqs}

By introducing \(\tilde{\rho}_{\vec{m}_L,\vec{C}}\) as the following:
\begin{equation}
\mathcal{E}^x_{p=1/2}(\hat{Z}^{\vec{m}_L} \rho_{\mathrm{RS}} \hat{Z}^{\vec{m}_L}) = \frac{1}{2^{|\vec{C}|}} \sum_{\vec{C}} |\vec{C}\rangle\langle\vec{C}| \otimes \tilde{\rho}_{\vec{m}_L,\vec{C}},
\end{equation}
We can write a term of Eq.~\eqref{Eq:F_channel_asym} as:
\begin{equation}
\mathcal{E}^x_{p=1/2}(\hat{Z}^{\vec{m}_L} \rho_{\mathrm{RS}} \hat{Z}^{\vec{m}_L} \hat{X}^{\vec{n}_{LR}}) = \frac{1}{2^{|\vec{C}|}} \sum_{\vec{C}} (-1)^{f(\vec{C})}|\vec{C}\rangle\langle\vec{C}| \otimes \tilde{\rho}_{\vec{m}_L,\vec{C}},
\label{eq:F_basic}
\end{equation}
The factor \( (-1)^{f(\vec{C})} \) in the sum corresponds to the sign of \(\hat{X}^{\vec{n}_{LR}}\) after the measurements. 

The feedback channel acts as:
\begin{equation}
F(\cdot) = \Tr_C \left( \sum_{\vec{C}} U_{\vec{C}} P_{\vec{C}} (\cdot) P_{\vec{C}} U_{\vec{C}}^\dagger \right),
\label{eq:F_channel}
\end{equation}
so we obtain:
\begin{align}
F\left( \mathcal{E}^x_{p=1/2}(\hat{Z}^{\vec{m}_L} \rho_{\mathrm{RS}} \hat{Z}^{\vec{m}_L} \hat{X}^{\vec{n}_{LR}}) \right) 
&= \frac{1}{2^{|\vec{C}|}} \sum_{\vec{C}} (-1)^{f(\vec{C})} \Tr_C \left( |\vec{C}\rangle\langle\vec{C}| \right) \tilde{\rho}_{\vec{m}_L} \nonumber \\
&= \frac{1}{2^{|\vec{C}|}}\left( \sum_{\vec{C}} (-1)^{f(\vec{C})} \right) \tilde{\rho}_{\vec{m}_L} = 0,
\label{eq:F_vanish}
\end{align}
Since \(\hat{X}^{\vec{n}_{LR}}\) is not a detector cell itself, the sign can be either positive or negative depending on the measurement outcome. This results in vanishing of this term when summing over \(\vec{C}\). Therefore, the feedback channel annihilates all non-symmetric terms.

We conclude:
\begin{equation}
F \left( \mathcal{E}^x_{p=1/2} \left( \mathcal{E}^z_{\theta}(\rho_{\mathrm{RS}}) \right) \right) 
= F \left( \mathcal{E}^x_{p=1/2} \left( \mathcal{E}^z_{p}(\rho_{\mathrm{RS}}) \right) \right),
\label{eq:final_result}
\end{equation}
for \(p = \sin^2{\alpha}\), establishing that coherent errors become effectively incoherent under the feedback channel. Therefore we have:
\begin{eqs}
    \mathcal{G} = F^{-1} \circ F
\end{eqs}

\end{document}